\documentclass[12pt]{article}

\pdfoutput=1

\usepackage{graphicx}
\usepackage{epstopdf}
\usepackage[centertags]{amsmath}
\usepackage{amssymb}
\usepackage{amsthm}
\usepackage{amsfonts}
\usepackage{ccaption}
\usepackage[usenames]{color}
\usepackage{mathrsfs}
\usepackage[
      colorlinks=true,
      linkcolor=blue,
      urlcolor=blue,
      filecolor=blue,
      citecolor=red,
      pdfstartview=FitV,
      pdftitle={},
        pdfauthor={Veronika Hubeny, Don Marolf, Mukund Rangamani},
        pdfsubject={Black holes on the brane},
        pdfkeywords={black holes, Hawking radiation, deconfinement},
        pdfpagemode=None,
        bookmarksopen=true
      ]{hyperref}

\usepackage{rotating}

\makeatletter
\@addtoreset{equation}{section}

\makeatletter
\renewcommand\section{\@startsection {section}{1}{\z@}%
                                   {-3.5ex \@plus -1ex \@minus -.2ex}
                                   {2.3ex \@plus.2ex}%
                                   {\normalfont\large\bfseries}}
\renewcommand\subsection{\@startsection{subsection}{2}{\z@}%
                                     {-3.25ex\@plus -1ex \@minus -.2ex}%
                                     {1.5ex \@plus .2ex}%
                                     {\normalfont\bfseries}}

  \captionnamefont{\bfseries}
  \captiontitlefont{\small\sffamily}
  \captiondelim{: }
  \hangcaption

\newcommand{\be}{\begin{equation}}
\newcommand{\ee}{\end{equation}}
\newcommand{\beq}{\begin{eqnarray}}
\newcommand{\eeq}{\end{eqnarray}}

\def\sec#1{\S\ref{#1}}
\def\fig#1{Fig.\,\ref{#1}}
\def\req#1{(\ref{#1})}

\def\({\left(}
\def\){\right)}
\def\[{\left[}
\def\]{\right]}
\def\vev#1{\langle\, #1 \, \rangle}

\def\ord#1{\CO\left(#1\right)}

\def\p{\partial}

\def\r{\rho}

\def\CB{{\cal B}}

\def\CD{{ \cal D }}

\def\CM{{\cal M}}
\def\CN{{\cal N}}
\def\CO{{\cal O}}

\def\CS{{\cal S}}
\def\CT{{\cal T}}

\def\ZZ{\mathbb{Z}}

\def\R{{\mathbb R}}
\def\Sp{{\bf S}}

\def\A5S5{{\rm AdS}_5 \times \S^5}

\def\p{\partial}

\def\ord#1{\CO\left( #1 \right)}

\def\vev#1{\langle\, #1 \, \rangle}

\definecolor{orange}{rgb}{0.8,0.3,0.8}

\def\AdS#1{AdS$_{#1}$}
\def\SAdS#1{Schwarzschild-AdS$_{#1}$}

\def\scri{\mathscr I}
\def\esu#1{\text{ESU}$_{#1}$}

\title{\bf \Large Hawking radiation from AdS black holes}

\author{\normalsize
Veronika E. Hubeny$^{a}$\footnote{veronika.hubeny@durham.ac.uk}, Donald Marolf$^{\,b\,}$\footnote{marolf@physics.ucsb.edu}, and
Mukund Rangamani$^{a}$\footnote{mukund.rangamani@durham.ac.uk} \\ \\
$^a$\small \sl  Centre for Particle Theory \& Department of
Mathematical Sciences,\\[-1.5mm]
\small \sl Science Laboratories, South Road, Durham DH1 3LE, United Kingdom. \\
$^b$ \small \sl Physics Department, UCSB, Santa Barbara, CA 93106, USA.
}
\begin{document}

\setlength{\baselineskip}{16pt}
\begin{titlepage}
\maketitle

\begin{picture}(0,0)(0,0)
\put(350, 320){DCPT-09/77}
\put(350,303){NSF-KITP-09-204}
\end{picture}
\vspace{-36pt}

\begin{abstract}
We study Hartle-Hawking-like states of quantum field theories on asymptotically AdS black hole backgrounds, with particular regard to the phase structure of interacting theories.   By a suitable analytic continuation we show that the equilibrium dynamics of field theories on large asymptotically AdS black holes can be related to the low temperature states of the same field theory on the AdS soliton (or pure AdS) background. This allows us to gain insight into Hartle-Hawking-like states on large-radius Schwarzschild- or rotating-AdS black holes. Furthermore,  we exploit the AdS/CFT correspondence to explore the physics of strongly coupled large $N$ theories on asymptotically AdS black holes. In particular, we exhibit a plausibly complete set of phases for the M2-brane  world-volume superconformal field theory on a BTZ black hole background.
Our analysis partially resolves puzzles previously raised in connection with Hawking radiation on large AdS black holes.
\end{abstract}
\thispagestyle{empty}
\setcounter{page}{0}
\end{titlepage}

\renewcommand{\thefootnote}{\arabic{footnote}}


\tableofcontents

\section{Introduction}
\label{s:intro}

Quantum fields on black hole spacetimes display a variety of physical phenomena, including vacuum polarization, particle production etc., see e.g.\ \cite{Birrell:1982ix} for a detailed discussion of these features.  However, even for free fields the details of such effects are typically difficult to calculate; for instance the expectation value of the stress tensor of a free conformally coupled scalar in the Schwarzschild black hole background has been computed analytically only in a WKB approximation \cite{Page:1982fm}. The situation is of course much worse for interacting theories.

The one universal feature of quantum fields on black hole spacetimes is that equilibrium states are thermal.  Indeed, for any field theory one can formally define the Hartle-Hawking state via a Euclidean path integral and, due to periodicity in Euclidean time, it will satisfy the Kubo-Martin-Schwinger (KMS) condition; see e.g.\ \cite{Jacobson:1995ak,Ross:2005sc} for reviews and references.   On asymptotically flat backgrounds, one may use the fact that this construction reduces to the thermal path integral in flat space in the region far from the black hole to understand properties of this equilibrium state.   For example, the stress tensor in this asymptotic region will be thermal at the black hole temperature.  Of course, non-equilibrium processes are more difficult to study; cf., \cite{Carter:1976di,FH,MacGibbon:1990zk,MacGibbon:1991tj,Hubeny:2009hr}.

However, not all black holes of interest are asymptotically flat.  Indeed,
asymptotically anti-de Sitter (AdS) black holes provide an important theoretical laboratory for studying many aspects of black hole physics. Historically, this was due to the fact that large \SAdS{} black holes have positive specific heat and are thermodynamically stable.  In addition, such black holes have recently been of great interest in the context of the Anti-de Sitter/conformal field theory (AdS/CFT) correspondence \cite{Maldacena:1997re} and its applications.  For such black holes, the fact that Anti-de Sitter space acts like a confining box means that physics in the asymptotic region remains coupled to the black hole.  As a result, despite the fact that Hartle-Hawking states again satisfy a KMS condition, interesting physical properties cannot be read off directly, even in the asymptotic region.

Our purpose below is to provide some tools to address this situation, and to better understand the physics of Hartle-Hawking-like equilibrium states of field theories on both static and rotating asymptotically AdS black hole backgrounds.  We are in particular motivated by a puzzle raised in \cite{Gregory:2008br}, which considered ${\cal N} = 4$ $SU(N)$ super-Yang-Mills (SYM) theory at large $N$ and strong coupling on \SAdS{4} backgrounds.  On large-radius \SAdS{4} black holes, \cite{Gregory:2008br} described a state where the stress tensor\footnote{Apart from a contribution due to the conformal anomaly which is proportional to the background metric.} was only ${\cal O}(1)$ at large $N$, and which was argued to dominate the thermodynamics.  The particular arguments given in \cite{Gregory:2008br} involved the \AdS{5}/CFT${}_4$ correspondence, but this fact will play no role in our main discussion.  The puzzle was that, while this theory is known to have an ${\cal O}(1)$ density of states at {\em low} temperatures on, say, the pure \AdS{4} background (confinement), it has an ${\cal O}(N^2)$ density of states (deconfinement) on \AdS{4} at high temperatures, above the so-called deconfinement transition \cite{Chamblin:2004vr}.  Intuitively this is because the background curvature of the \AdS{} spacetime provides an IR cut-off for the field theory, effectively putting it in a box of size $\ell$, the \AdS{} radius. (This makes it similar to the theory on $\Sp^3 \times \R$ where one has a confinement-deconfinement transition at large $N$ \cite{Witten:1998zw}.) Thus, due to the high Hawking temperature, one might naively expect similar deconfined behavior on large-radius \SAdS{4} backgrounds and a corresponding ${\cal O}(N^2)$ stress tensor, contrary to actual findings.

In this paper, we address this puzzle, and the general issue of characterizing Hartle-Hawking-like states of asymptotically AdS black holes, in several stages.  We first review some basic properties of \SAdS{} black holes in \sec{s:prelim}, where we also discuss various possible notions of confinement/deconfinement for interacting theories on asymptotically AdS backgrounds.  Due to the fact that AdS acts like a finite box, these notions are somewhat more subtle than on asymptotically flat backgrounds.

We then argue in \sec{s:ftads} that a particular analytic continuation (specifically a double Wick rotation) can give a useful perspective on Hartle-Hawking-like states of asymptotically AdS black holes; we consider both static and rotating black holes.  For the special cases of the 3-dimensional Ba\~nados-Teitelboim-Zanelli (BTZ) black holes, this technique provides an exact map between rotationally-invariant equilibrium states on BTZ backgrounds satisfying a certain ${\mathbb Z}_2$ symmetry and similar such states on pure \AdS{3} for any quantum field theory.  The interesting point is that this transformation maps states on {\em high} temperature BTZ backgrounds to {\em low} temperature states on \AdS{3} (and vice versa).  In higher dimensions, a similar precise map can be found between $U(1)$-invariant equilibrium states (again satisfying a ${\mathbb Z}_2$ symmetry) on planar-\AdS{d} black hole backgrounds and such states on \AdS{d} soliton backgrounds.\footnote{The AdS soliton background as we review later is related by a double Wick rotation to the planar \SAdS{} black hole.} Since large-radius \SAdS{4} black holes are well approximated by the planar \AdS{4} black hole geometry, these results in
particular resolve the puzzle raised in  \cite{Gregory:2008br} (at least partially) if the theory `confines'  in the sense of having an ${\cal O}(1)$ vacuum energy on the \AdS{4} soliton background.

While the arguments of section \sec{s:ftads} hold for any quantum field theory, they only suffice to relate phases on the above pairs of backgrounds.  More input is needed to determine the actual phases that arise for any particular theory.  We therefore turn in \sec{s:phases} to a detailed investigation of the phases of a particular class of theories related to the puzzle of \cite{Gregory:2008br}, namely the 3- and 4-dimensional strongly interacting large $N$ theories which admit an AdS/CFT correspondence.

We begin with the ${\cal N}=4$ SYM theory studied in \cite{Gregory:2008br} (or more generally a large $N$ field theory with an \AdS{5} holographic dual).  We find evidence that the theory does indeed confine with ${\cal O}(1)$ vacuum energy on the \AdS{4} soliton. As noted above and  discussed in greater detail in \sec{s:ftads}, this partially resolves the puzzle described in \cite{Gregory:2008br}.  We also show that other mechanisms  which lead to small stress tensors such as those described in \cite{Fitzpatrick:2006cd,Hubeny:2009hr} do not apply to this case.  In particular, we find no sign of confinement in the prescribed phase on the original \SAdS{4} background itself.   However, finding the classifying set of phases for $d$ dimensional CFTs on asymptotically AdS backgrounds with $d > 3$ remains an interesting open problem.

To better understand the phase structure of similar theories on AdS black holes, we turn  our attention to low dimensional CFTs in \sec{s:phases}. Specifically,  we also investigate three-dimensional conformal field theories arising on the M2-brane world-volume \cite{Seiberg:1997ax} or generalizations thereof, (cf.\ \cite{Aharony:2008ug} for recent developments in understanding these theories) on BTZ black hole backgrounds. In this case, we are able to construct a (plausibly) complete set of dual gravity solutions by exploiting the conformal flatness of the BTZ black hole.  The dual bulk geometries are asymptotically \AdS{4} geometries whose boundary metric is conformal to BTZ. The relevant dual geometries turn out to be the \AdS{4} bubble of nothing or the so-called BTZ black string spacetime. The former spacetime is the dominant saddle point for low temperature BTZ black holes, while the latter holographically captures the physics  on high temperature BTZ black hole backgrounds.  This analysis therefore allows us to identify relevant phases of the theory  for BTZ black holes of any temperature and angular velocity.  The general picture emerging from this detailed analysis again shows that an $\CO(1)$ stress tensor in a given phase need not be correlated with other properties familiar from confined phases in flat space. We close with a brief discussion in \sec{s:discuss}.

\section{Thermodynamics and confinement on asymptotically AdS backgrounds}
\label{s:prelim}

We commence our discussion of quantum fields on non-dynamical asymptotically AdS spacetimes, by noting some of the key features that are special to these geometries.  We will mainly focus on asymptotically AdS black hole geometries, though we will have occasion to discuss some non-black hole spacetimes as well. The key point distinguishing asymptotically AdS spacetimes is of course the timelike nature of $\scri$, which in particular implies that one needs to specify additional boundary conditions in order to discuss classical and quantum fields in these geometries.\footnote{In this paper we will extensively use the fact that the  AdS spacetimes are conformal to half of an Einstein static universe. This in particular means that we can inherit boundary conditions from the ESU for quantum fields in AdS, a choice which sometimes is referred to as `transparent boundary conditions'. We will employ these boundary conditions exclusively in the current discussion. A more general discussion of other boundary conditions will appear elsewhere.} 
  Relatedly, the global AdS geometry acts as a confining box, regulating some of the long-distance IR divergences that usually plague discussions of thermal physics in asymptotically flat spacetimes. In fact, as originally discussed in \cite{Callan:1989em} one could use AdS as a natural infra-red regulator to study  interacting quantum field theories;\footnote{We would like to thank David Tong for bringing this to our attention and for useful discussions on this issue.} we will have occasion to discuss this in detail below.

\subsection{Thermal behaviour of AdS black holes}
\label{s:prelimA}

It is natural to begin with static spherically symmetric black holes in \AdS{d} spacetimes, whose metric is given by
\begin{equation}
ds^2 = -f_{d}(r) \, dt^2 + \frac{dr^2}{f_{d}(r)} + r^2\, d\Omega_{d-2}^2 \ ,
\label{sadsbh}
\end{equation}
with
\begin{equation}
f_d(r) = \frac{r^2}{\ell^2} +1 - \left(\frac{r_+}{r}\right)^{d-3}\, \left( \frac{r_+^2}{\ell^2}+1 \right) \ .
\label{fdr}
\end{equation}
where $\ell$ determines the AdS scale and $r_+$ is the horizon radius.
While small \AdS{d} black holes (those with $r_+ \ll  \sqrt{\frac{d-3}{d-1}}\, \ell$)
are similar to their asymptotically flat counterparts, the thermodynamic properties of large AdS black holes are quite different.  For example, large AdS black holes have positive specific heat and are thermodynamically stable, while asymptotically flat black holes are thermodynamically unstable. This difference results from the interaction between the horizon and the asymptotically AdS boundary conditions and has much to do with the diverging redshift experienced by observers near AdS infinity. This fact plays an important role in the AdS/CFT correspondence where the large black holes in \AdS{d} are dual to the high temperature phase of the dual field theory.

The temperature of a \SAdS{d} black hole is
\begin{equation}
T_H = \frac{(d-1) \, r_+^2 + (d-3)\, \ell^2}{4 \pi\, r_+ \, \ell^2} \ .
\label{bhtemp}
\end{equation}
However, from the point of view of local dynamics in these spacetimes, the diverging redshift near the boundary ensures that the high temperatures associated with large values of $r_+$ are not locally observable.  In particular, for a free field in the Hartle-Hawking state outside a large \SAdS{} black hole, no static observer ever experiences a local temperature significantly greater than the Unruh temperature associated with their proper acceleration, despite the fact that the black hole temperature $T_H$ diverges as $r_+ \to \infty$.  While the state is not precisely thermal with respect to freely falling observers, it is nevertheless true that no local freely falling observer anywhere outside the horizon would find significant excitations above the AdS scale; see for example \cite{Brynjolfsson:2008uc}.  This fact may come as a surprise to readers most used to thinking about asymptotically AdS spacetimes as the AdS (or bulk) side of the AdS/CFT correspondence.  However, one  should recall the key role played in AdS/CFT by the conformal rescaling performed to obtain field theory observables in the latter context. By stripping off the leading $r^2$ fall-off in the gravitational potential, the AdS/CFT dictionary relates local results in the CFT to the global AdS temperature \req{bhtemp}.

A simple argument that local AdS measurements do not see high temperatures is  to recall that large \SAdS{d} black holes are well-approximated by so-called planar AdS black holes with translationally invariant horizons.  In particular, for $r_+ \gg \ell$, \req{sadsbh} with \req{fdr} reduces to
\begin{equation}
ds^2 = -\frac{r^2}{\ell^2}\,\left(1- \frac{r_+^{d-1}}{r^{d-1}}\right) \, dt^2 + \frac{
\, dr^2}{\frac{r^2}{\ell^2}\,\left(1- \frac{r_+^{d-1}}{r^{d-1}}\right)} + r^2\, d{\bf x}_{d-2}^2 \ .
\label{planarbh}
\end{equation}
Now any two such planar black holes are related by a diffeomorphism (see e.g.\ \cite{Horowitz:1998ha}), since the parameter $r_+$  can be absorbed by scaling the coordinates $t, {\bf x}$. This means that local observations are independent of the temperature $T = \frac{(d-1) \, r_+}{4\pi\, \ell^2}$ of the planar black hole and, since the AdS scale is the only scale in the problem, local observers see no excitations above this scale. This conclusion was verified in \cite{Hemming:2007yq} using a simplified model based on spherical reduction, and in \cite{Gregory:2008br} using the Page approximation \cite{Page:1982fm} for a conformally coupled scalar field.  Thus, while from a global perspective large \SAdS{} black holes become very hot, locally one might say that they are at most only lukewarm. In what follows we will often use $T_H$ to parameterize the black hole, but in light of the above we refrain from using the terms `hot'  or `cold' to describe the associated ensemble.

Given that large AdS black holes never get really hot from the local perspective, one might wonder if this result could be enough to resolve the puzzle raised in \cite{Gregory:2008br} regarding the small stress tensor on
large \SAdS{} black holes. Recall that \cite{Gregory:2008br} considered a large $N$ conformal theory with $N^2$ degrees of freedom, such as the $SU(N)$ $\CN =4$ SYM on asymptotically AdS spacetimes. These theories have a density of  states of $\ord{N^2}$ at large temperatures on \AdS{4} (which is similar to their behaviour on $\R^{3,1}$). However, a holographic computation\footnote{As we review in \sec{s:phases}, the holographic dual spacetime is itself a known negatively curved AdS black string geometry in one higher dimension, whose asymptotic behavior readily yields the corresponding boundary stress tensor.} revealed that the theory actually has only an $\ord{1}$ stress tensor in the Hartle-Hawking state on high-temperature \SAdS{4} backgrounds.  In \cite{Gregory:2008br} this was interpreted as evidence for a confined phase (in parallel with the suggestion of \cite{Fitzpatrick:2006cd}).  The surprise was that confinement would occur at high temperatures. Given that local observers do not experience high temperatures, we see that there is no immediate tension.  Indeed, the theory in question is known to have an ${\cal O}(1)$ density of states on pure  global \AdS{4} at temperatures below the confinement transition at $T_c \sim \ell^{-1}$, where $\ell$ is the \AdS{} scale.   However, our detailed investigation of this phase in \sec{s:sads} will show that other expected properties of confinement fail to hold in this phase.    Instead, we will suggest an alternate mechanism leading to small  i.e. ${\cal O}(1)$, stress tensors.\footnote{In fact as we describe later the situation is reminiscent of $\CN =4$ SYM on a Scherk-Schwarz circle, where naively the true vacuum state has an $\CO(N^2)$ stress tensor, while an excited state has $\CO(1)$ stress tensor at zero temperature. In this case the resolution is simple: one just measures physical quantities relative to the true vacuum  state.}

To prepare for the above mentioned analysis, we must first address what is meant by confinement on asymptotically AdS backgrounds, especially in the presence of black holes.  A moment's thought reveals that this notion is not as straightforward as one might expect. We now turn to discuss the issue in some detail.

\subsection{Notions of confinement in quantum gauge theories}
\label{s:confnotions}

In interacting field theories, phase transitions that change the number of low-energy effective degrees of freedom are commonplace.  As a result, the response of the field theory to long-wavelength disturbances, such as deformations of the background metric, can differ greatly between the two sides of the transition.  Of particular interest to us are the confinement-deconfinement transitions, present in certain gauge theories.  To understand the distinction between the phases we will recall several established notions of confinement which are available and describe their relative merits for applications to quantum fields on our backgrounds. The upshot of our discussion will be that while the last of the definitions listed below is more appropriate than the other two for horizon-free asymptotically AdS spacetimes, no conventional definition provides a clean notion of confinement in asymptotically AdS black hole backgrounds (suggesting that perhaps no such definition exists).

\paragraph{1. The $q{\bar q}$ potential:} For non-abelian gauge theories such as QCD in flat space, confined phases can be defined in terms of the induced potential between external quarks and anti-quarks. In a confining theory the external quark can be introduced only at an infinite energy cost.  Correspondingly, the $q{\bar q}$  potential diverges linearly at long distance, or equivalently the expectation value of the spatial Wilson loop exhibits an area law.  In this context, one can also define a confinement scale associated with the distance at which the potential transitions from some (generally softer) short-distance behavior to the long-distance linear potential.

\paragraph{2. Mass gap:} An alternative viewpoint on confinement is to phrase the question in terms of the dynamical generation of a mass gap in theories with no massive fundamental matter. For instance in large $N$ pure QCD in flat space, due to the non-trivial renormalization group flow, one expects that the theory dynamically generates a  mass scale $\Lambda_\text{QCD}$, though proving this of course still remains an unsolved problem.

\paragraph{3. The density of states:} Yet another way to characterize the physics of confinement is to consider the density of states $\CD(E)$ which measures the number of available states in some small energy band centered at $E$. This is expected to be small in the confined phase: for instance, due to the diverging long-distance potential, quarks are bound into a relatively small number of color-neutral excitations such as hadrons or glueballs.  Now if we consider a $SU(N)$ gauge theory at large $N$, the number of such singlet states is only ${\cal O}(1)$ while the number of charged excitations would have been ${\cal O}(N^2)$.  It therefore follows that in the confined phase the entropy is\footnote{A more invariant characterization of this phenomenon is in terms of the central charge $c$ at the UV fixed point of the theory; in the confined phase the entropy is $\ord{1}$, while it is $\ord{c}$ in the deconfined phase.}   $S = \log \CD(E) \sim \ord{1}$, as is the density of states.
Of course, on Minkowski space the actual density of states diverges in either phase, but the density of states per unit volume is well-defined and behaves as above.

While these notions are inter-related in certain familiar circumstances, it is useful to pick one definition for use below.  In order to do so, we now briefly discuss their relative merits in the context of current interest.

\subsection{Confinement on asymptotically AdS backgrounds}
\label{s:adsconf}

Characterizing the confined and deconfined phases directly in terms of the density of states is quite useful when we consider gauge theories on  spatially compact manifolds, such as $\Sp^3 \times {\mathbb R}$, where the notion of a long-distance potential is ill-defined on scales $L$ larger than the size $R$ of the $\Sp^3$; see e.g.\ \cite{Aharony:2003sx}. Field theories on Anti-de Sitter space appear to be another such context where the long-distance $q\bar q$ potential fails to provide a useful characterization of confinement  (see \cite{Callan:1989em} for an early discussion) and the density of states definition is preferred.  Firstly, we note that any theory in AdS experiences a diverging gravitational potential at infinity, though this has little to do with the separation of charges.
Secondly, we wish to examine the dynamics of conformal field theories such as $\CN =4$ SYM on these spacetimes, to make contact with the issues raised in \cite{Gregory:2008br}.

To elaborate on this second point, recall that \AdS{d} is conformally equivalent to (half of) the Einstein static universe, \esu{d} $ = \Sp^{d-1} \times {\mathbb R}$, and conformal gauge theories confine on
the \esu{}. The latter statement is in fact a consequence of kinematics; gauge invariance (through the Gauss law constraint) on compact spatial manifolds forces the low energy excitations to be singlets. One expects therefore that,  with an appropriate choice of boundary conditions, such theories should also confine on \AdS{d}.  However, this confinement will not be associated with a divergence in the $q{\bar q}$ potential (other than the diverging gravitational potential noted above).  The point is that the conformal transformation maps the equator of \esu{d} to the asymptotic region of \AdS{d} and, as a result, relates the long-distance behavior in \AdS{d} directly to short-distance properties on  the \esu{}.  This is basically controlled by the short-distance UV fixed point of the theory and should be universal for all states. In fact one finds that the non-gravitational part of the potential has a finite limit at large \AdS{d} separations in all phases (confined or deconfined) of the theory.  While it is still interesting to ask if the potential in some phase may display linear behaviour over some range of scales $L_{max} \gg L \gg L_{min}$, we will not take this as a fundamental definition of confinement.

As a matter of fact, an early discussion of confinement in AdS spacetimes can be found in \cite{Callan:1989em} where the authors argue that the negative curvature of the background provides a natural setting to study infra-red physics of interacting field theories such as QCD. Noting that the conventional Wilson loop area law (or the static $q{\bar q}$ potential) doesn't capture the essence of confinement (for reasons described above or more heuristically by noting that the negative curvature implies that areas and volumes scale the same way), the authors of \cite{Callan:1989em} propose looking at annular Wilson loops separated in the AdS radial direction, i.e., imagining that the $q {\bar q}$ pair are situated at $r = r_q$ and $r = r_q + \delta_q$ respectively with $r_q \gg \ell$. One would then obtain a small subleading contribution to the $q{\bar q}$ potential on top of the divergent result (for $r_q \gg \ell$) due to the AdS asymptotics. However, since this notion of confinement is a small effect buried underneath the curvature effects, one expects it to correlate only weakly with other properties of the state such as the expectation value of the stress tensor. In addition, since this criterion is only sensitive to the asymptotic behavior of the state, for conformal field theories it is again determined entirely by UV properties of the vacuum on the \esu{}.

On the other hand, because the gravitational potential of pure \AdS{} effectively puts the theory in a ``box'' of size set by the AdS scale $\ell$, the density of states is finite at finite $N$.  Its large $N$ scaling thus gives a sharp criterion to define confined versus deconfined phases which is not determined by a purely UV effect.  We will therefore  use this property below to define confinement in horizon-free asymptotically AdS spacetimes.  We leave any relation between the density of states and a possible linear regime in the $q\bar q$ potential as a topic for future investigation.

While this takes care of horizon-free backgrounds, the most interesting asymptotically AdS backgrounds will contain black holes.  In such cases, the extreme redshift at the black hole horizon typically introduces new infra-red (IR) divergences.  For free fields one can readily show that the Hamiltonian has continuous spectrum, and one expects a similar result for any well-defined quantum field theory.  As a result, comparing the density of states in two phases would require some notion of an IR cutoff near the horizon. Of course, this also makes the notion of the mass gap hard to use to characterize confinement. 

While one could seek a useful construction of such a cut-off, and while one could attempt to compare various phases in the limit where this cutoff is removed, one does not expect the result to be particularly convenient as a classification tool.  This is because black hole horizons can be polarized by nearby charges, even if the black hole has no net charge.   In particular, if one envisions charged particle in a confined phase as being connected by string-like flux tubes, the flux tubes from charges close to the horizon can effectively end on the horizon instead of on other particles.  As a result, separating pairs of charges in the direction along the black hole horizon requires little energy, so long as the charges are sufficiently close to the horizon. One therefore expects that the density of states behaves in the same (deconfined) manner in any state as this IR cut-off is removed, no matter what the local physics farther from the black hole is.  As a result,  we will generally avoid speaking of confined or deconfined phases on black hole backgrounds.  We will also avoid discussing the density of states on these backgrounds, though as advertized we will investigate other features (such as the $q{\bar q}$ potential) related to the above concepts of confinement in \sec{s:phases}.

\section{Field theories on AdS black holes}
\label{s:ftads}

Having described some basic issues concerning quantum field theory and phase transitions on asymptotically AdS black hole backgrounds, we now turn to a more detailed discussion of quantum fields in these backgrounds.  We begin with the general structure of equilibrium states on stationary AdS black hole backgrounds. While we could focus directly on so-called global black holes such as the \SAdS{} spacetime \req{sadsbh}, it is convenient to first consider the planar black holes \req{planarbh} which approximate
\req{sadsbh} in the large black hole regime.  This has the advantage of not only being a bit simpler, but also of drawing out the key features that distinguish AdS black holes from their flat space counterparts. Furthermore, to facilitate explicit computations, it is useful to consider the three dimensional BTZ black hole spacetime, which again offers useful insights.  We will however keep both the mass and the angular momentum of these BTZ black holes as free parameters, in order to explore the full phase space of such stationary configurations.

Our primary interest is in equilibrium states for the quantum fields (although we allow for unstable phases), for which the notion of equilibrium implies a time-translation symmetry that allows for analytic continuation.  We also assume a second spatial symmetry corresponding to translations for planar \AdS{d} and rotations in the global case and for BTZ.\footnote{Readers puzzled by our assertion that equilibrium Hartle-Hawking states exist for rotating AdS black holes (in contrast to their asymptotically flat cousins for which such states do not exist \cite{Kay:1988mu}) should see footnote \ref{rotHH} in \sec{s:ftplanar} for an explanation.}
  Furthermore, we will demand that simultaneous time and parity reversal along the spatial isometry direction leave the quantum state invariant. This discrete $\ZZ_2$ symmetry allows us to consider double Wick rotations, which we will use as a crutch to identify some of the features of interest. We begin with a discussion of quantum fields in BTZ  spacetimes in \sec{s:ftbtz} and then turn to the higher dimensional examples in \sec{s:ftplanar}.

\subsection{Field theories on BTZ black hole background}
\label{s:ftbtz}

Let us first discuss field theories on BTZ black holes.  The (rotating) BTZ spacetime is described by the metric
\begin{equation}
ds^2 = -\frac{(r^2-r_+^2)(r^2-r_-^2)}{\ell^2\, r^2} \, dt^2 + \frac{r^2\, \ell^2}{(r^2-r_+^2)(r^2-r_-^2)}\, dr^2 + r^2\, \left(d\phi - \frac{r_+\,r_-}{\ell\,r^2}\, dt\right)^2
\label{BTZmet},
\end{equation}
where $\ell$ is the AdS scale and $r_\pm$ parameterize the outer and inner horizons of this stationary spacetime, in terms of which one can express the mass $M$ and angular momentum $J$ parameters\footnote{Note that when the coordinate $\phi$ is allowed to range over entire ${\mathbb R}$,  the metrics \req{BTZmet} for all $r_\pm$ are diffeomorphic to each other and to pure \AdS{3} \cite{Banados:1992gq}.  However, the parameters $M, J$ take on invariant meanings if one defines $\phi$ to be periodic.  We take this period to be $2\pi$ so that $M$ and $J$ are the usual mass and angular momentum of the BTZ black hole.
}  as
\begin{equation}
M = \frac{1}{\ell^2}\, \left(r_+^2 + r_-^2\right) , \qquad J  = \frac{2}{\ell} \, r_+\, r_- \ .
\label{mjTBZ}
\end{equation}

Since we are interested in Hartle-Hawking-like states, it is natural to consider the complexified BTZ spacetimes. Recall that if we consider the static BTZ black hole with $r_- =0$ then the analytically continued Euclidean geometry can be used to define the Hartle-Hawking state of the quantum field theory since it naturally implements the KMS condition. The more general stationary spacetimes \req{BTZmet} are periodic under the combined transformation
\begin{equation}
(t,\phi) \rightarrow (t + \frac{i}{T _\text{BTZ}} , \phi + \frac{i \,\Omega}{T_\text{BTZ}})
\end{equation}
 for
\begin{equation}
T_\text{BTZ}= \frac{r_+^2 - r_-^2}{2\pi r_+\, \ell^2 }, \qquad  \Omega = \frac{r_-}{ r_+\, \ell},
\label{tempb}
\end{equation}
as well as the original $(t, \phi) \rightarrow (t , \phi + 2 \pi)$. Since any choice of $T_\text{BTZ}$ and  $|\Omega\, \ell| \le 1$ determines a unique $M$, $|J| \le M \ell$,  thermodynamic quantities on this family of backgrounds are functions of $T_\text{BTZ}$ and  $\Omega$; e.g., the partition function is $Z = Z\left(T_\text{BTZ}, \Omega\right)$.

The key point for us is that each complexified solution above also admits a second interpretation.  Specifically, the transformation
\begin{equation}
\tilde t = -i  \( {r_- \over \ell} \, t - r_+ \, \phi \) \ , \qquad
\tilde \phi = -i  \( {r_+ \over \ell^2} \, t - {r_- \over \ell} \, \phi \) \ , \qquad
\tilde r = \ell \, \sqrt{{r^2 - r_+^2 \over r_+^2 - r_-^2}}
\label{}
\label{BTZcoordchange}
\end{equation}	
maps the BTZ spacetime \req{BTZmet} to global \AdS{3}
%
\begin{equation}
ds^2 = - \left(  \frac{{\tilde r}^2}{\ell^2} +1 \right) \, d{\tilde t}^2
+ \frac{d{\tilde r}^2}{\left(  \frac{{\tilde r}^2}{\ell^2} +1 \right) }
+ {\tilde r}^2 \, d {\tilde \phi}^2
\label{}
\end{equation}	
%
with periodicities:
\begin{equation}
\label{BTZperiods}
(\tilde t, \tilde \phi) \rightarrow (\tilde t, \tilde \phi + 2 \pi)  \quad {\rm and} \quad  (\tilde t, \tilde \phi) \to
(\tilde t + \frac{i}{T_\text{AdS}} , \tilde \phi + \frac{i \, \Omega}{T_\text{AdS}} ), \quad {\rm for} \quad
T_\text{AdS} = \frac{{1 - \Omega^2\, \ell^2}}{4 \pi^2\,T_\text{BTZ}\, \ell^2}.
\end{equation}
Thus the complexified BTZ metric also represents thermal \AdS{3} at the new temperature $T_\text{AdS}$ and the original chemical potential $\Omega$.   (It is for this reason that we do not label $\Omega$ with a subscript.)  It follows that the properties of Hartle-Hawking-like states on BTZ are given by analytic continuation of those on pure \AdS{3} with the mapping of temperatures given by \req{BTZperiods}; for instance
\begin{equation}
Z_\text{BTZ}(T_\text{BTZ}, \Omega) = Z_\text{AdS}(T_\text{AdS}=\frac{{1 - \Omega^2\, \ell^2}}{4 \pi^2\,T_\text{BTZ}\, \ell^2} , \Omega) \ .
\label{pfnrel}
\end{equation}
 In particular, since in the limit $T_\text{BTZ} \to \infty$ any thermodynamic quantity on BTZ is determined by the \AdS{3} phase at $T_\text{AdS} =0$, if one uses the \AdS{3} vacuum as a reference point\footnote{This reference point is a natural one, since, by the maximal symmetry of \AdS{3}, any non-zero answer can be regarded as the renormalization of some background parameter.  For example, any non-zero \AdS{3}-invariant stress tensor can be regarded as a renormalization of the cosmological constant.} it follows that any thermodynamic quantity will vanish as $T_\text{BTZ} \rightarrow \infty$.

There is a geometric way to explain the relation between the BTZ and thermal AdS geometries as originally described in \cite{Maldacena:1998bw}. The point is that one can view the Euclidean BTZ or thermal AdS spacetimes as hyperbolic three-manifolds whose boundary is a two-torus $T^2$. Depending on which one-cycle of the $T^2$ shrinks to zero in the interior of the spacetime, we obtain a specific member of the two-parameter family of solutions \req{BTZmet} (after appropriate analytic continuation).  In this language the choice of the shrinking cycle is captured by the complex structure parameter of the $T^2$  and the map between thermal AdS and  BTZ is just a modular transformation of this  complex structure.

The above arguments apply for any field theory. In order to make contact with known results, it is useful to first discuss free theories. Since \AdS{3} acts like a confining box of size $\ell$, the low energy states of any free field theory will have energies of order $\omega \sim \ell^{-1}$ so that their contributions to the partition function (or any other thermodynamic quantity) will be exponentially suppressed for $T_\text{AdS} \, \ell \ll 1$.  In particular, the expected stress tensor will be exponentially small at high $T_{BTZ}$.

This is precisely the behavior found by explicit calculation in \cite{Steif:1993zv} for the expected stress tensor $\langle \CT_\nu^\mu \rangle_{HH}$ of a conformally-coupled scalar in the Hartle-Hawking state on the BTZ background. There it was shown that, for the static black hole with $r_- =0$ (or equivalently $\Omega =0$),
 \begin{equation}
 \vev{\CT^\mu_\nu}_{HH} = \frac{A(r_+)}{r^3} \, {\rm diag} \{1,1,-2\}   \quad {\rm for} \quad A(r_+) =
  \frac{2}{32\, \pi} \, \sum_{n=1}^\infty\, \frac{\cosh 2\pi\, n\, r_+ + 3}{(\cosh 2\pi\, n\, r_+ -1)^{3/2}}\ ,
 \label{steifJ0}
\end{equation}
which indeed vanishes exponentially for $T_\text{BTZ} \sim r_+ \gg 1$.  The  stress tensor for free fields in the general case with $r_- \neq 0$ is described in \cite{Steif:1993zv}; we will describe the analogous result for strongly coupled quantum fields  in \sec{s:kerrbtz}.

One may also reproduce the results of \cite{Steif:1993zv} at small $T_\text{BTZ}$ by using the analytic continuation described above and the fact that \AdS{3} can be mapped conformally into \esu{3}.   Applying these operations to the
stress tensor
\begin{equation}
\label{freeESUstress}
\vev{\CT^\mu_\nu}_{\Sp^2 \times \R} = \sigma_{3} \, T^3 {\rm diag}\{-2,1,1\}
\end{equation}
on $\Sp^2 \times \mathbb{R}$ yields a result that agrees with \req{steifJ0} for small $r_+$.   In particular, the interchange of $t$ and  $\phi$ under the double Wick rotation for $\Omega =0$  explains why the stress tensor found in \cite{Steif:1993zv} has a negative energy density (and is in fact proportional to ${\rm diag}\{1,1,-2\}$).
Note that taking $\sigma_{3}$ to be the 2+1-dimensional Stefan-Boltzman constant, i.e., $\sigma_3 = 2\, \zeta(3)$ the stress tensor \req{freeESUstress} describes the thermal state of any free scalar on $\Sp^2 \times \mathbb{R}$ at large $T$.

Let us now consider interacting theories with non-trivial phase structure, where \req{pfnrel} implies that for a given value of $\Omega$, the high temperature phases on BTZ are in direct correspondence with the {\em low} temperature phases on AdS. One generally expects that there is a unique such low temperature phase, typified by the ground state.  Furthermore, in a large $N$ confining gauge theory one expects that for small $T_\text{AdS}$ (below the phase transition), thermodynamic quantities receive contributions only from the ${\cal O}(1)$ singlet degrees of freedom even at large $N$.  In such theories, we see that thermodynamic quantities will be independent of $N$ at large $T_\text{BTZ}$.

\subsection{Field theories on planar AdS black hole backgrounds}
\label{s:ftplanar}

The simplest examples of higher dimensional asymptotically AdS black hole spacetimes are the planar AdS black holes with translationally invariant horizons \req{planarbh}. We will also consider field theories on boosted planar \AdS{d} black holes, for they can be studied by precisely the same methods as in \sec{s:ftbtz} and describe the large-radius limit of rotating AdS black holes.  These spacetimes can be obtained from \req{planarbh} by a simple relabeling  $x_{d-2} = \ell\,\phi$ and a coordinate transformation which is a boost with rapidity $\alpha$ in the $(t,\phi)$ plane, resulting in the metric
\begin{equation}
ds_{planar}^2 =  \frac{r^2}{\ell^2}  \left( - dt^2 + \ell^2\,  d \phi^2 + \frac{r_+^{d-1}}{r^{d-1}} (\cosh \alpha \ dt - \ell \sinh \alpha \ d \phi)^2 \right) + r^2\, d{\bf x}_{d-3}^2 + \frac{\ell^2 \,dr^2}{r^2\,  \(1 -\frac{r_+^{d-1}}{r^{d-1}}\)}
\label{poinbh}
\end{equation}	
for $d > 3$, where again we use $\ell$ to denote the AdS length scale.
When the coordinate $\phi$ is allowed to range over all of ${\mathbb R}$, the metrics (\ref{poinbh}) are diffeomorphic for all $r_+$  and $\alpha$.  However, the parameters $r_+$ and $\alpha$ take on invariant meanings if one takes $\phi$ to be periodic (say, with period $2\pi$).  In that case, choosing a conformal frame in which the length of the $\phi$ circle is also $2 \pi$ in the boundary metric, the energy and momentum densities are
\begin{equation}
\frac{M}{V_{d-3}} = \frac{1}{8\, G_{d}^{(N)}} \, \cosh^2 \alpha  \, \(\frac{(d-1)\,r_+}{4}\)^{\! d-1}, \quad \frac{P}{V_{d-3}} =  \frac{1}{8\, G_{d}^{(N)}}\, \sinh \alpha \cosh \alpha \(\frac{(d-1)\,r_+}{4}\)^{\! d-1} .
\label{stplanar}
\end{equation}

As before, we will be interested in complexified spacetimes in order to understand the Hartle-Hawking states.
Apart from the usual $(t, \phi) \rightarrow (t , \phi + 2 \pi)$, these are again periodic under the transformation
\begin{equation}
(t,\phi) \rightarrow \(t + \frac{i}{T_\text{planar}} , \ \phi + \frac{i \,\Omega}{T_\text{planar}}\) ,
\label{}
\end{equation}
 for
\begin{equation}
T_\text{planar} =  \frac{(d-1)\, r_+}{4 \pi\, \ell^2 \,\cosh \alpha}\ ,  \qquad \Omega = { \tanh \alpha \over \ell}\ ,
\label{Tplanar}
\end{equation}
which of course captures the basic fact that the boosted planar black hole \req{poinbh} corresponds to a grand canonical ensemble at temperature $T_\text{planar}$ and momentum chemical potential $\Omega$.

Let us now consider a complex coordinate transformation
\begin{eqnarray}
\left(
        \begin{array}{c}
		\tilde t \\
		\tilde \phi \\
        \end{array}
\right)
&=& - i\, \frac{2 \pi \, T_\text{planar} \, \ell^2}{1-\Omega^2\, \ell^2}
 \left(
	\begin{array}{cccc}
		 \Omega & - 1 \\
                {1 \over \ell^2}  & - \Omega \\
	\end{array}
\right)
\left(
	\begin{array}{c}
 	 t \\
	\phi \\
	\end{array}
\right) ,\nonumber \\
\tilde{r} &=& \frac{\sqrt{1-\Omega^2\, \ell^2}}{2\pi\, T_\text{planar}\,\ell } \, r \ , \qquad
\tilde{{\bf x}} = \frac{2\pi\, T_\text{planar}\,\ell}{\sqrt{1-\Omega^2\, \ell^2}} \;{\bf x}\ ,
\label{pl2sol}
\end{eqnarray}
which transforms the metric (\ref{poinbh}) to another familiar solution, the \AdS{d} soliton \cite{Horowitz:1998ha}:
\begin{equation}
ds_{soliton}^2 =  \frac{{\tilde r}^2}{\ell^2}  \, \left(- d{\tilde t}^2 + \(1-\frac{{\tilde r}_+^{d-1}}{{\tilde r}^{d-1}}\) \ell^2\,  d {\tilde \phi}^2  \right) + {\tilde r}^2\, d{\tilde {\bf x}}_{d-3}^2 + \frac{\ell^2 \,d{\tilde r}^2}{{\tilde r}^2\,  \(1 -\frac{{\tilde r}_+^{d-1}}{{\tilde r}^{d-1}}\)}
\label{adssol}
\end{equation}
where ${\tilde r}_+ = \tilde r(r_+)$ as in \req{pl2sol}.  The rescalings employed in \req{pl2sol} are
in fact the unique ones
such that the spatial circle of the soliton has period $ 2\pi$. It is now easy to check that the coordinates in \req{adssol} are identified as
\begin{eqnarray}
&& (\tilde t, \tilde \phi)  \rightarrow (\tilde t , \tilde \phi + 2\pi) \quad{\rm  and} \quad (\tilde t, \tilde \phi) \rightarrow
\(\tilde t + \frac{i}{T_\text{sol}}  , \ \tilde \phi + \frac{ i\, \Omega}{T_\text{sol}} \), \nonumber \\
&& \quad {\rm for } \quad
T_\text{sol} = \frac{{1 - \Omega^2\, \ell^2}}{4 \pi^2\,T_\text{planar}\, \ell^2} \ .
\end{eqnarray}

Thus we learn that the complexification of \req{poinbh} also represents the \AdS{d} soliton, with the spatial circle having period $2 \pi$ at a different temperature $T_\text{sol}$  and the original chemical potential $\Omega$. A different way to say this statement is to realize that there is a complex metric which has two different real Lorentzian sections: one analytic continuation leads to the planar black hole, while the other leads to a grand-canonical ensemble for the AdS soliton.\footnote{The complex geometry which captures the saddle point has a (complex) $T^2$ boundary (the $(t,\phi)$ plane) and the mapping to the AdS soliton is once again a modular transformation of the complex structure of this torus.}

As a result of this analytic continuation,  field theories on planar \AdS{d} black holes behave similarly to those on BTZ as discussed earlier in \sec{s:ftbtz}. The high-temperature behavior on the planar black hole is determined by the low-temperature behavior of the field theory on the \AdS{d} soliton with the spatial circle having fixed period ($2 \pi$), since the two gravitational solutions are related by an analytic transformation. In particular, we learn that the high $T_\text{planar}$ limit of the stress tensor will be finite and will just be the analytic continuation of the vacuum stress tensor on the \AdS{d} soliton background.

The main difference from the BTZ case is that for $d >2$ the AdS soliton differs from \AdS{d} and this limiting stress tensor need not vanish.  However, temperature-dependent corrections will still come from the low-energy modes on the AdS soliton.  For free field theories, one may expand in modes in the $\phi, r$ directions and express the result as a $(d-2)$-dimensional theory (in the $\tilde{t}, {\tilde x}$ directions) featuring a tower of massive fields, with the lightest mass being roughly the AdS scale $\ell^{-1}$.  So, in parallel with the BTZ case, temperature-dependent corrections are exponentially suppressed in $(T_\text{sol}\,  \ell)^{-1} \propto T_\text{planar} \, \ell$.

For interacting theories with phase transitions, the high $T_\text{planar}$ phase of the field theory on the black hole is related to the (presumably unique) phase at low $T_\text{sol}$ on the AdS soliton.  In particular, let us consider a large $N$ gauge theory which confines in flat space below some energy scale $\Lambda$. If $\Lambda \gg \ell^{-1}$, the large $T_\text{planar}$ stress tensor should be independent of $N$ (i.e., ${\cal O}(1)$) above some critical temperature $T_\text{planar}^*$. For cases with $\Lambda \lesssim \ell$, the large $T_\text{planar}$ stress tensor is again related by analytic continuation to behavior at small $T_\text{sol}$. Since we are then at low temperatures $T_\text{sol}$ in a confined phase, one expects that all thermal corrections remain independent of $N$ for sufficiently large $T_\text{planar}$.  However, since the confinement and curvature scales are comparable, there might now be a non-trivial zero point energy on the soliton which scales with $N$, so that the limiting large $T_\text{planar}$ stress tensor does as well.\footnote{To illustrate the point about non-trivial zero point energy, consider $\CN=4$ SYM on a $\R^{2,1} \times \Sp^1_{ss}$ where we impose anti-periodic boundary conditions for fermions on $\Sp^1_{ss}$. While there is a unique vacuum state of this theory which confines, due to the fact that at low energies one essentially has a 2+1 dimensional pure glue theory \cite{Witten:1998zw}, explicit computations using holographic techniques described in \sec{s:phases} shows that one has an $\CO(N^2)$ vacuum energy in this confining vacuum.}   We will see explicit examples of this type of situation below.

As already indicated, the above analysis can  be used as a starting point to understand Hartle-Hawking states on large Schwarzschild- or rotating-AdS black hole backgrounds.  One should be able to treat the states on these black holes with large but finite radius as mild deformations of states on planar black holes or, under double Wick rotation, of thermal states on AdS solitons.\footnote{\label{rotHH}The rotating AdS black holes are expected to have a Hartle-Hawking vacuum \cite{Hawking:1998kw} (see also \cite{Emparan:2008eg} for a recent discussion) which corresponds to thermal equilibrium as long as $\Omega \, \ell \leq 1$, in contrast to the situation for asymptotically flat rotating black holes. In the latter case, the existence of superradiant modes makes the Hartle-Hawking state singular \cite{Kay:1988mu}. However, for $\Omega \, \ell \leq 1$
with asymptotically AdS boundary conditions,
there is a Killing field (the horizon generator) which is timelike everywhere outside the horizon.  This leads to a positive definite conserved quantity which forbids unstable modes and renders the Hartle-Hawking state well-defined.}

This provides a partial resolution to the puzzle posed in \cite{Gregory:2008br}, where it was found that the stress tensor of four dimensional ${\cal N} =4$ SYM theory at large $N$ was only ${\cal O}(1)$ on large-radius \SAdS{4} backgrounds.  This field theory is known to confine when at least one dimension is compactified with supersymmetry breaking boundary conditions \cite{Witten:1998zw},
or subjected to some other infra-red cutoff, for example, on $\Sp^3 \times {\mathbb R}$, and therefore on \AdS{4} as defined by a conformal transformation.  As a result, it is natural to expect the theory to confine on the \AdS{4} soliton as well.\footnote{Here it is important to realize that one has to impose anti-periodic boundary conditions for the fermions along the spatial circle of the soliton, for the periodic boundary condition is not compatible with the circle being contractible.} Thus, as above, one expects purely thermal corrections at small $T_\text{sol}$ to be independent of $N$, and this confinement may also provide a mechanism to suppress corrections due to the mild deformation\footnote{The deformation is small for large $T_\text{SAdS}$.  However, since the deformation appears to involve changes at the AdS scale $\ell$ which sets the IR cutoff inducing confinement, it remains surprising that the corrections are so strongly suppressed.} required to take the \AdS{4} soliton to double Wick rotations of \SAdS{4}.  Now, as with the $\Lambda \lesssim \ell$ case discussed above, there remains the possibility of a large ground state energy on the \AdS{4} soliton.  However, we will use AdS/CFT to directly argue  in section \sec{s:sads} that such a large ground state energy does not arise, and also to confirm the above statements.

\subsection{Interlude: Other mechanisms for suppressing stress tensors}
\label{s:compare}

We have seen that the BTZ and planar \AdS{d} black holes can be related to pure \AdS{3} and the \AdS{d} soliton respectively, which led us to argue that thermal corrections to the stress tensor of weakly coupled large $N$ quantum gauge theory on these backgrounds is $\ord{1}$ for large $T_\text{BTZ}$ and $T_\text{planar}$. In addition, we have  argued that the BTZ stress tensor vanishes in the limit of large $T_\text{BTZ}$.  It is interesting to compare our discussion above with other mechanisms that can lead to small stress tensors for large $N$ theories on high-temperature black hole backgrounds.

One such mechanism was exhibited in \cite{Fitzpatrick:2006cd} in the context of asymptotically flat black holes. The idea is that, although the thermodynamically dominant phase at large $T$ in flat space is deconfined, there is also a meta-stable confined phase. Suppose that the state far away from the black hole is initially confined.  Then at large $N$ a large amount of heat is required to induce the deconfinement transition.  Although the black hole acts as a heat source, the confined phase conducts heat from the black hole only slowly.  As a result, at least in the limit of large $N$, such states can perhaps remain in approximate equilibrium for arbitrarily long times.  Under such circumstances, one expects an ${\cal O}(N^2)$ stress tensor near the black hole, but which becomes small far away.

We recently described another such mechanism in \cite{Hubeny:2009hr}. There
we identified cases where the physical size of excitations in the deconfined plasma appeared to be much larger than the size of the black hole.  This resulted in weak coupling between black hole and plasma, producing states in which the stress tensor, even if large, decreases rapidly far from the black hole.   This mechanism could function together with that of \cite{Fitzpatrick:2006cd}, though it can  also act in cases without a well-defined confinement transition.  In particular, we argued that such situations were possible even for conformal theories such as ${\cal N}=4$ SYM.

The present mechanism of confinement in the double Wick rotated background is rather different in several ways.  Firstly, the mechanisms of \cite{Fitzpatrick:2006cd} and \cite{Hubeny:2009hr} tend to produce ${\cal O}(N^2)$ stress tensors, which however fall off far from the black hole.  In contrast, confinement in the double Wick rotated background means that the response of the stress tensor to certain deformations is small everywhere in the spacetime.  Secondly, this mechanism does not require the theory to confine in any sense away from the black hole, nor the thermal plasma to couple weakly.  Indeed, we analyze some example phases in \sec{s:phases} below which demonstrate that neither of these effects need be associated with confinement in the double Wick rotated background.

\section{Strongly coupled field theories on AdS black holes}
\label{s:phases}

Having described some general tools for studying quantum fields on asymptotically AdS black hole backgrounds in \sec{s:ftads}, we now investigate the dynamics of certain strongly coupled field theories on these backgrounds in detail. Our main interest is in large $N$ conformal field theories of the type associated with the puzzle of \cite{Gregory:2008br}.  In particular, we will consider $\CN =4$ SYM  theory studied in \cite{Gregory:2008br} and its relatives\footnote{Here we have in mind the $\CN =1$ super-conformal quiver gauge theories which are constructed in terms of D3-branes probing various singularities.}, as well as the 3-dimensional theory of M2-branes.  Such theories can be investigated at strong coupling using the AdS/CFT correspondence.

Our primary goal is to confirm the general expectations for such theories described at the end of \sec{s:ftplanar}. Since the analysis of \sec{s:ftads} holds for any theory, we can map high temperature phases on BTZ black holes to low temperature phases on pure \AdS{3}, and similarly in higher dimensions we can map high temperature phases on planar AdS  black holes to low temperature phases on the AdS soliton. We then use the AdS/CFT correspondence to investigate phases on such static backgrounds and argue that the theories confine for low $T_{\text{AdS}}$, $T_\text{sol}$ respectively.  We also contrast the behavior on the black hole backgrounds with the mechanisms for suppressing stress tenors described in \cite{Fitzpatrick:2006cd,Hubeny:2009hr}.  For the simpler case of the M2-brane superconformal field theory at large $N$ we construct an apparently complete set of phases for all $(T_\text{BTZ}, \Omega)$, including the low $T_\text{BTZ}$ regime which is difficult to study in the higher-dimensional case.  Interestingly, we find that the low $T_\text{BTZ}$ phase displays a linear $q{\bar q}$ potential at appropriate values of $r$, though the stress tensor is $\CO(N^\frac{3}{2})$.

\subsection{Prelude: A framework to explore strong coupling}
\label{s:prelude}

As stated above, we will use the holographic AdS/CFT correspondence \cite{Maldacena:1997re,Gubser:1998bc,Witten:1998qj} to investigate dynamics. While we will not review such techniques in detail here, the essential idea is that in the limit of large $N$ and large 't Hooft-like coupling states of certain field theories on given $d$-dimensional spacetimes $\CB_d$ are dual to solutions of supergravity theories that asymptotically approach \AdS{d+1}$\times X$, where $X$ is a compact manifold, whose isometries (if any) are realized as global symmetries of the field theory.  Below, both $\CB_d$ and the holographic dual will be asymptotically AdS spacetimes.  To avoid confusion, we refer to $\CB_d$ as the field theory spacetime or the field theory black hole, distinguishing it from the higher dimensional `bulk' spacetime which may also contain a (bulk) black hole.

Though the methods of \sec{s:ftads} can also be used for any stationary axisymmetric phase, we consider phases below which preserve the full set of field theory global symmetries.  This restriction suppresses dynamics in the internal space $X$.  As a result, it is natural to look for the associated bulk solutions in the sector where the bulk spacetime takes the form $\CM_{d+1} \times X$ for an appropriate $X$ , where $\CM_{d+1}$ satisfies the Einstein equations with negative cosmological constant $\Lambda = -\frac{d\,(d-1)}{2}\,L_{d+1}^{-2}$ as described by the action\footnote{To avoid confusion we will use $L_{d+1}$ to denote the bulk \AdS{d+1} length
scale of the holographic dual, reserving $\ell_d$ for the AdS length scale on the field theory background $\CB_d$.}
\begin{equation}
\CS_{\text{bulk}}  = \frac{1}{16 \pi \, G^{(d+1)}_N}\,\int d^{d+1}x \, \sqrt{-g}\, \left( R - 2 \, \Lambda\right)  \ .
\label{bulkact}
\end{equation}	
This captures the universal behavior of a large class of field theories.
To describe the theory on the particular background $\CB_d$, we require that $\CM_{d+1}$ admit a conformal compactification with conformal boundary $\CB_d$.
The $(d+1)$-dimensional Newton's constant is related to the central charge of the conformal field theory via:
\begin{equation}
c = \frac{L_{d+1}^{d-1}}{16\pi\, G_N^{(d+1)} }  \ .
\label{centralch}
\end{equation}
Of course, $d=4$ for theories defined by D3-branes and $d=3$ for M2-brane theories.

\subsection{Four dimensional SCFTs on large \SAdS{4} black holes}
\label{s:sads}

We now specialize to four-dimensional large $N$ conformal field theories obtained from D3-branes, which we place on asymptotically \AdS{4} backgrounds. Our aim will be to construct 5-dimensional spacetimes $\CM_5$ whose conformal boundary is the prescribed asymptotically AdS spacetime $\CB_4$ with metric $\gamma_{\mu\nu}$.  Let us begin with the following observation. Given an asymptotically \AdS{d} Einstein metric on $\CB_d$ with cosmological constant set by the curvature scale $\ell_d$, we can immediately construct one particular solution to the bulk equations of motion \req{bulkact}. Specifically, we can simply warp $\CB_d$ in an extra bulk-radial direction $R$,  as in  \cite{DeWolfe:1999cp}.  Then the $d+1$ dimensional metric
\begin{equation}
ds^2 = dR^2 + \frac{L_{d+1}^2}{\ell_d^2} \, \cosh^2\(\frac{R}{L_{d+1}}\)\, \gamma_{\mu \nu} \, dx^\mu\, dx^\nu  = \frac{L_{d+1}^2}{\cos^2\Theta} \, \(d\Theta^2 + \frac{1}{\ell_d^2} \, \gamma_{\mu\nu}\, dx^\mu\,dx^\nu \)
\label{adsinads}
\end{equation}
solves the equations of motion arising from \req{bulkact} as long as $\gamma_{\mu\nu}$ satisfies the same equations in one lower dimension (and with cosmological constant set by $\ell_d$, i.e.\ $\Lambda_\text{bdy} = -\frac{(d-1)(d-2)}{2\, \ell_d^2}$).

As noted in \cite{Gregory:2008br}, this result implies that the so-called \SAdS{} black string spacetime in $d+1$ dimensions is a bulk solution describing the AdS/CFT dual of a rotationally-invariant equilibrium state of CFTs on \SAdS{d}.   The explicit bulk  \SAdS{d+1} black string solution with \SAdS{d} boundary metric takes the form \req{adsinads} with $\gamma_{\mu\nu}$
given by \req{sadsbh} i.e.,
\begin{equation}
ds^2 = dR^2  + \frac{L_{d+1}^2}{\ell_d^2} \, \cosh^2\(\frac{R}{L_{d+1}}\)\,
\(-f_{d}(r) \, dt^2 + \frac{dr^2}{f_{d}(r)} + r^2\, d\Omega_{d-2}^2 \) \ .
\label{sadsbs}
\end{equation}
with $f_d(r)$ as in \req{fdr}.  The large $r_+$ limit yields a similar black string constructed using the planar black hole \req{planarbh} instead of \req{fdr}. Now, generically the black string geometries \req{sadsbs} could suffer from Gregory-Laflamme  type instablities. However, for large $r_+$ it can be shown that the \SAdS{} black string is stable to linearized gravitational perturbations \cite{Hirayama:2001bi}, suggesting that it describes the equilibrium Hartle-Hawking state of the field theory.\footnote{Note that this is in contrast with the AdS black strings of \cite{Chamblin:1999by} with the boundary metric being asymptotically flat.  There, the only scale is given by the boundary black hole size $r_+$, so the string is unstable for any $r_+$ \cite{Gregory:2000gf}; from the bulk point of view this is because the string becomes arbitrarily thin near the Poincare horizon.  On the other hand, in the present situation of interest, the minimal bulk thickness attained by our Schwarzschild-AdS black string is given by $r_+$, rendering the stability for $r_+ > L_{d+1}$ natural. The analysis of \cite{Hirayama:2001bi} also demonstrates explicitly that the \SAdS{} black string is unstable when $r_+ \leq \ell_{d}$ as one would expect in analogy with Schwarzschild black strings in Kaluza-Klein geometries.}

However, any state dual to a bulk metric of the form \req{adsinads} has an ${\cal O}(1)$ field theory stress tensor at large $N$.\footnote{One can obtain the stress tensor for the geometries \req{adsinads} using the counter-term method outlined in \cite{Balasubramanian:1999re} or using holographic renormalization techniques \cite{deHaro:2000xn}.} This follows directly from the analysis of \cite{Gregory:2008br} for the \SAdS{} black string and one arrives at the same conclusion for more general boundary metrics by noting that the form of \req{adsinads} constrains the fall-off of the bulk metric asymptotically as $R \to \infty$ (or equivalently as $\Theta \to \frac{\pi}{2}$).  The basic construction clearly generalizes to rotating \AdS{d} black holes as well, and one expects similar stability results at large radius so long as $|\Omega\, \ell_{d}| < 1$.\footnote{In $d> 4$ rotating \AdS{d} black holes can carry angular momenta in different rotating planes and one requires that $|\Omega_i \,\ell_d | \leq 1$ for all these rotations.}

To understand the situation better,  in parallel with the discussion of \sec{s:ftplanar}, we consider analytic continuations of the large $r_+$ \SAdS{} black string. It is clear that one need only apply these operations to the metric $\gamma_{\mu\nu}$ on $\CB_d$ and then insert the result into \req{adsinads}.  The double Wick rotation of \req{sadsbh} is the AdS bubble of nothing whose large $r_+$ limit is simply the AdS soliton of \cite{Horowitz:1998ha}.

Let us therefore consider the ``AdS soliton-string'' metric\footnote{This solution was recently examined in the braneworld context in \cite{Chen:2008vh}.} given by taking $\gamma_{\mu\nu}$ in \req{adsinads} to be the \AdS{d} soliton \req{adssol}.  The resulting $d+1$ spacetime is static and horizon-free.  The metric has a non-compact translational symmetry ${\tilde x}_k \to  {\tilde x}_k + \xi_{{\tilde x}_k}$ for $k = 1, \cdots , d-2$  with no fixed points, and a $U(1)$ symmetry $\tilde \phi \to \tilde \phi + 2 \pi$ which acts trivially on the surface ${\tilde r} = \text{constant}$.  Furthermore, the gravitational potential grows steeply in the $R$ and ${\tilde r}$ directions, for in each of these we attain an AdS asymptopia.  As a result, metric perturbations are described by a $(d-2)+1$ dimensional effective theory with mass scale $m \sim \ell_d^{-1}$.  While the density of states diverges due to the non-compact translational symmetry in the $\R^{d-2}$, the density of states per unit ${\bf {\tilde x}}$-length is finite and  moreover is of ${\cal O}(1)$.   Thus, for $d=4$ this solution is dual to a confining phase of the large $N$ SCFT (e.g., the  $\CN =4$ SYM) on the zero-temperature \AdS{4}-soliton background as previously predicted.  In making this final statement, one is making the reasonable assumption that the zero-temperature phase is unique.  The generalization to rotating black holes is also straightforward\footnote{In the rotating case, it is natural here to Wick rotate the time direction $t$ and the azimuthal direction $\phi$.  The result is a locally stationary axi-symmetric bubble of nothing spacetime with observer-dependent horizons, which again approaches the 5-dimensional AdS soliton \req{adssol} in a certain large-radius limit.  These solutions differ somewhat from the AdS-Kerr bubbles obtained in \cite{Balasubramanian:2002am}, which follows \cite{Aharony:2002cx} in Wick rotating the polar angle $\theta$ instead of the azimuthal angle $\phi$.  Of course, these two analytic continuations define the same bubble spacetime (in different coordinates) for static, spherically symmetric black holes.} since in the large $r_+$ regime we can use the boosted planar black hole discussed in \sec{s:ftplanar}.

It is also clear by the usual arguments that this phase describes a `linear $q{\bar q}$ potential'.  In particular, consider a timelike Wilson loop in the field theory on the \AdS{4} soliton \req{adssol} at some fixed value of ${\tilde r}$, say ${\tilde r} = {\tilde r}_+$ and  separated in the single spatial direction ${\tilde x}$ by an amount large compared with the period of $\tilde \phi$.  In the bulk AdS soliton-string spacetime i.e., in \req{adsinads} with the metric $\gamma_{\mu\nu}$ taken to be the \AdS{4} soliton metric \req{adssol}, the expectation value of this  Wilson loop is dual to the (regulated) area of the string world-sheet anchored on this Wilson loop.
The piece of string along this surface gives rise to the linear potential at large separations in ${\tilde x}$.  Thus, as usual in horizon-free static spacetimes with simple asymptotics, here the two notions of confinement agree.

In contrast, there are no signs of confinement for the five dimensional \SAdS{} black string.  Instead, the density of states diverges due to the presence of a horizon. In addition, Wilson lines with large separations are dual to string world-sheets that sink down close to the bulk horizon and which therefore cost very little energy to separate.

This behavior is clearly very different from the confined-state physics associated with the mechanism described in \cite{Fitzpatrick:2006cd} for obtaining small stress tensors on high temperature black holes.   It is also very different from the mechanism described in \cite{Hubeny:2009hr}, which stems from an unusually weak coupling of the field theory plasma to the field theory black hole.  As discussed in \cite{Hubeny:2009hr}, the dual description of such a weakly coupled setting is a so-called black droplet horizon.  Such horizons cap off smoothly in the bulk and thus allow bulk gravitons (or even hypothetical massive bulk particles) to pass ``under'' the horizon without interacting strongly.  In contrast,  the 5-dimensional \SAdS{} black-string horizon continues all the way through the bulk, ending only on a second asymptotically AdS boundary on the opposite side.

In fact, the \SAdS{} black string has rather more in common with the so-called black funnel solutions which were also introduced in \cite{Hubeny:2009hr}.  Such solutions were argued to provide bulk duals to Hartle-Hawking states for which the field theory plasma couples strongly to the black hole.  The definition of black funnels given in \cite{Hubeny:2009hr} was limited to asymptotically flat boundaries $\CB_d$ (and closely related spacetimes), in which context they are characterized by having a single connected horizon which connects the boundary black hole in $\CB_d$ with an asymptotic region dual to a deconfined plasma far from the field theory black hole.

However, at least in the presence of an appropriate Killing field, we can offer a better way to characterize the difference between the droplet and funnel horizons which does not rely on $\CB_d$ having particular asymptotic behavior. Suppose that $\CM_{d+1}$ has a rotational Killing field which also induces a rotational Killing field on $\CB_d$.  Now consider bulk geodesics which are `boundary-radial' in the sense of having vanishing angular momentum with respect to this isometry.
This class of geodesics is most relevant to our considerations because such trajectories most closely mimic plasma excitations in the boundary theory  `aimed at' the black hole.  We may then define horizons connected to the boundary black hole
to be black funnels if all boundary-radial geodesics cross the horizon, while they are black droplets if there is an open set of boundary-radial geodesics which avoid falling through the horizon.  This revised definition coincides with that of \cite{Hubeny:2009hr} when $\CB_d$ has a rotational Killing field. It also coincides with that of \cite{Hubeny:2009ta} for the cases studied there (which in general had no well-defined asymptotics for $\CB_d$).  In the above sense, both \SAdS{}- and rotating-AdS black strings are clearly black funnels in the sense that all excitations couple strongly to the field theory black hole.

This completes our analysis of $\CN=4$ SYM and its cousins on \SAdS{} backgrounds.  As argued in \cite{Gregory:2008br}, the stability of this solution for large $T_\text{SAdS}$ suggests that it describes the dominant high-temperature phase.  However, the instability at low $T_\text{SAdS}$ argues that some other phase dominates in this regime.  We were unable to determine the nature of this phase, nor could we determine if there are additional bulk solutions which give rise to further phases of the boundary CFT on high temperature \AdS{d} black holes. One could attempt to find a more general class of solutions numerically, but we now turn to the simpler case $d=3$ where we believe that we can construct a full set of bulk geometries. For this simple case, we will see that the field theory has a unique phase on high temperature black hole backgrounds, and that this phase is again described holographically by \SAdS{} (or rotating-AdS) black strings.

\subsection{Three dimensional SCFTs on  BTZ black hole background}
\label{s:btz}

We argued in \sec{s:ftads} that the phases of a field theory on \AdS{d} black holes can be related to the phases on either pure \AdS{3} (for $d=3$) or the AdS soliton (in $d>3$) by an appropriate analytic continuation. We have also seen above that while the holographic AdS/CFT correspondence is useful to elucidate some of the properties of CFTs on AdS black hole backgrounds, we are somewhat crippled in being unable to determine the full phase structure owing to lack of knowledge regarding the complete set of solutions with the given boundary conditions.  In what follows we will focus on the $d=3$ case, where we believe can find all the relevant four dimensional bulk geometries $\CM_4$ which are holographically dual to strongly coupled CFTs on the BTZ background.

The crucial fact that we exploit to construct the relevant geometries $\CM_4$ for $\CB_3 = \text{BTZ}$ is the
observation that by a suitable conformal transformation one can map  the (complexified) BTZ spacetime to a known boundary metric.  First of all  we recall that  \AdS{3} with length scale $\ell$ is conformally related to (half of) the Einstein-static universe $\Sp^2 \times {\mathbb R}$ of radius $\ell$ and moreover under this conformal mapping one typically uses the same time coordinate on both \AdS{3} and $\Sp^2 \times {\mathbb R}$, and thus also the same notion of temperature.  As in \sec{s:ftbtz}, it is related to the temperature $T_\text{BTZ}$ of the associated BTZ black hole (with the same AdS length scale $\ell$) by \req{BTZperiods} when one transforms thermal \AdS{3} to BTZ by the appropriate Wick rotation. Second, the dynamics of CFTs on $\Sp^2 \times {\mathbb R}$ is well studied in the AdS/CFT literature and we can exploit this to understand the various possible phases of the field theory.

Before turning to the actual discussion from the bulk \AdS{4} viewpoint, we pause to recall some of the salient features about $2+1$ dimensional CFTs which are known to have holographic duals. A large class of such field theories can be constructed as the world-volume theories on a stack of $N$ M2-branes probing various singularities, see  \cite{Aharony:2008ug} for explicit description of such theories. The central charge of these theories in the holographic regime scales as $c \sim N^{\frac{3}{2}}$ and the holographic dual spacetimes are of the form \AdS{4}$\times X_7$ where $X_7$ is an Einstein manifold.\footnote{\label{M2c} By explicit computation using the M2-brane near horizon geometry we find the central charge for the $\CN =8$ M2-brane world-volume theory with $SO(8)$ R-symmetry to be $c = \left( \frac{2}{3} \right)^4 \pi^3 (2N)^{3/2} $. }

 Rotationally-invariant equilibrium states of these CFTs on   $\Sp^2 \times {\mathbb R}$ are dual to stationary axisymmetric bulk solutions.  Allowing for a chemical potential for rotation in the $\Sp^2$, for states preserving the full global  symmetry of the field theory (i.e., no variation in the internal space $X$),
 the relevant solutions are presumed to be those given by the rotating-\AdS{4} black holes \cite{Carter:1968cz} of mass $M$ and angular momentum $J$ (or equivalently by their temperature $T$ and angular velocity $\Omega$).  Here we explicitly include the pure \AdS{4} metric defined by the limit $M,\, J \to 0$. We begin in \sec{s:sadsbtz} by discussing the canonical ensemble where $J =0$, so that we are considering the M2-brane theory on the non-rotating BTZ black hole background given by \req{BTZmet} with $r_- =0$. Subsequently, we will analyze the more general situation allowing for non-zero $J$ in \sec{s:kerrbtz}.

In the following since we only discuss three dimensional boundary geometries ($d=3$) we will drop the subscripts on the bulk and boundary AdS scales $L$ and $\ell$ to keep the notation clean.

\subsubsection{Static BTZ black hole}
\label{s:sadsbtz}

To understand the behavior of 3-dimensional CFTs on the static BTZ spacetime, we will start with the known behavior of thermal CFTs on $\Sp^2 \times \R$. The relevant asymptotically \AdS{4} solutions which (plausibly) capture the phases holographically are the \SAdS{4} black holes parameterized by their mass $M$. As described above, this set of solutions can be mapped via appropriate analytic continuation to obtain the relevant bulk geometries when the boundary spacetime is BTZ of a given temperature $T_\text{BTZ}$.

While the transformations from $\Sp^2 \times \R$ to BTZ  can be applied directly to the bulk solution, we find it conceptually simpler to use dS$_{2} \times \Sp^1$ as an intermediate step, where dS$_2$ is $1+1$ de Sitter space. In particular, we will:
\begin{itemize}
\item Analytically continue the Einstein static universe $\Sp^2 \times \R$ to dS$_{2} \times \Sp^1$.
\item  Conformally map  dS$_{2} \times \Sp^1$ to BTZ.  This transformation maps the cosmological horizon of dS$_2$ into the black hole horizon of BTZ.
\end{itemize}

Let us therefore start with a static, spherically symmetric, asymptotically \AdS{4} spacetime in global coordinates:
\begin{equation}
ds^2 = - f(\rho) \, dT^2 + { d\r^2 \over f(\rho)}
+ \rho^2 \( d\theta^2 + \sin^2 \theta \, d\Phi^2 \)
\label{SAdS4}
\end{equation}	
for some $f(\rho)$ which asymptotes as $\rho \to \infty $ to $ \frac{\rho^2}{L^2} + 1 -\ord{\rho^{-1}}$.
A double Wick rotation
\begin{equation}
T = i \, \chi \ , \qquad \Phi = i \, \tilde{t} \
\label{wick}
\end{equation}	
then casts the metric \req{SAdS4} into the form
\begin{equation}
ds^2 =  f(\rho) \, d\chi^2 + { d\rho^2 \over f(\rho)}
+ \rho^2 \( - (1 - {\tilde r}^ { \, 2}) \, d{\tilde t}^{ \, 2} + {d {\tilde r}^ { \, 2} \over 1 - {\tilde r}^ { \, 2}} \)
\label{SAdS4met3}
\end{equation}	
where we have also performed a coordinate change  $\cos\theta = \tilde{r}$. This metric is an asymptotically \AdS{4} geometry whose boundary is now dS$_{2} \times \Sp^1$, with dS$_2$ size $L$ inherited from the \AdS{4} scale, and the $\Sp^1$ size determined from the original black hole temperature.

To further recast \req{SAdS4met3} into a form which has conformally BTZ boundary metric, we merely need to let
\begin{equation}
{\tilde t} = \frac{r_+}{\ell^2} \, t \ , \qquad
{\tilde r} = \frac{r_+}{r} \ , \qquad
\chi = \frac{L}{\ell} \, r_+ \, \phi  \ ,
\label{dStoBTZcx}
\end{equation}	
which yields the following bulk metric:
\begin{equation}
ds^2 = \frac{\rho^2 \, r_+^2}{r^2\, \ell^2} \,  \[ -
 \frac{r^2 - r_+^2}{\ell^2} \, \, dt^2 + {\ell^2 \over r^2 - r_+^2} \, dr^2
+  r^2 \,  {L^2 \, f(\r) \over \r^2} \, d\phi^2 \] +  { d\r^2 \over f(\r)} \ .
\label{sadsbtzF}
\end{equation}	
Since as $\rho \to \infty$, the factor in parenthesis is the standard BTZ metric with radius $r_+$ and AdS$_3$ scale $\ell$, it is clear that the boundary is conformal to BTZ. The rescaling of the $\chi$ coordinate used in \req{dStoBTZcx} was performed to ensure that the $\phi$ circle has period $2\pi$.  Assuming that after our analytic continuation the $\chi$ circle has period $\Delta \chi$, this identifies $r_+ = \frac{\ell}{L}\, \frac{\Delta \chi}{2\pi} $.
Note that both the pure \AdS{4} and the \SAdS{4} geometries may be written in the form \req{sadsbtzF}, and that these are the only static spherically symmetric asymptotically AdS solutions with $\Sp^2 \times {\mathbb R}$ boundary metric \cite{Chrusciel:2000az}.  This observation justifies the choice of the ansatz \req{SAdS4}.

For the special case where the initial seed metric was pure \AdS{4}, taking $f(\rho) = \frac{\rho^2}{L^2} +1$, by an appropriate change of coordinates\footnote{Specifically,
\begin{equation*}
\cosh^2 \frac{R}{L} = 1+ \frac{r_+^2\, \rho^2}{L^2 \, r^2} \ , \qquad
\hat{r}^2 = r_+^2\, \left(1+\frac{\rho^2}{L^2}\right)  \, \left(1+\frac{r_+^2\, \rho^2}{L^2 \, r^2}\right)^{\! -1}
\label{}
\end{equation*}	
with $\hat{r}$ being the radial coordinate for the BTZ part of the metric $\gamma_{\mu\nu}$ in \req{adsinads}.}
the result \req{sadsbtzF} can be written in the form \req{adsinads} with $\gamma_{\mu \nu}$ the static BTZ black hole of radius $r_+$; the result is just a BTZ black string.\footnote{Here we use the terminology of \cite{Emparan:1999fd}, where the solution was obtained as a special case of the AdS C-metric.}
One can equivalently obtain the solution (see \cite{Brill:1996zp,Aminneborg:1996iz,Banados:1997df}) by writing \AdS{4} in an \AdS{3} slicing and applying the quotient construction \cite{Banados:1992gq} that turns \AdS{3} into BTZ.  Such solutions are free of naked singularities for all choices of the period $\Delta \chi$, since the circle stays non-contractible everywhere in the spacetime; as a result we find solutions for all values of  $r_+$ or equivalently for all  $T_\text{BTZ}$, cf., \req{tempb}. These solutions  are also the 4-dimensional analogue of the asymptotically \AdS{5} \SAdS{} black strings studied in \cite{Gregory:2008br} and \sec{s:sads}.

In contrast, for the \SAdS{4} black hole where
\begin{equation}
f(\rho) =  \frac{\rho^2}{L^2} + 1 - \frac{\mu}{\rho}  \ , \qquad \mu =\rho_+  \left(1 + \frac{\rho_+^2}{L^2}\right),
\label{fsads4}
\end{equation}	
the circle is contractible, since the function $f(\rho)$ vanishes at $\rho= \rho_+$. It follows that $\chi$ is an angular variable and its  period fixes $r_+$ to be
\begin{equation}
r_+= \frac{\ell}{L} \, \frac{2}{f'(\rho)}\bigg|_{\rho =\rho_+}  = \frac{2\, \rho_+ \,  \ell\,  L}{3\, \rho_+^2 + L^2} \  ,
\label{dcsads4}
\end{equation}	
so that $T_{BTZ} = \frac{1}{4 \pi^2 \, \ell \, L \, T_{\text{Schw-AdS}_4}}$, where $T_{\text{Schw-AdS}_4}$ is the original temperature of the bulk \SAdS{4} black hole.  As one would expect, this relation differs from \req{BTZperiods} only by the factor of $\ell/L$ required to relate the bulk and boundary \AdS{} length scales.
As a result the \SAdS{4} geometries limit themselves to providing bulk duals for BTZ geometries with $r_+ \le \frac{\ell}{\sqrt{3}}$, i.e., only for small (and thus low temperature) BTZ black holes. Note also that this relation is double valued; there are two different values of $\rho_+ \in \R^+$ both of which lead to the same $r_+$. In the case of $\rho_+ \neq 0$,  \req{sadsbtzF} describes the
static region of the \AdS{4} bubble of nothing studied in \cite{Balasubramanian:2002am}.

In summary, we find that for $T_\text{BTZ} \leq T_\text{c}= \frac{1}{2\pi\,\sqrt{3}\, \ell}$; we have three possible solutions. The first is the BTZ black string, and the other two are the bubble solutions arising from the \SAdS{4} black hole. Since the Euclidean action of a bubble is the same as that of the corresponding \SAdS{4} black hole, we readily identify the larger bubble as a local minimum of the free energy and the smaller bubble as an unstable saddle point corresponding to a local maximum of the free energy.  Furthermore, there is a first order phase transition at  some $T_\text{BTZ} =  T_\star$.\footnote{The value of $T_\star$ can be easily computed by comparing the free energies of the solutions. One finds that the solutions exchange dominance at $\rho_+ = L$ which implies that $r_+ = \frac{\ell}{2}$ and thence $T_\star = \frac{1}{4\pi\, \ell}$. See also \fig{f:rotbtz}, \fig{f:rotbtzT}.}
The large bubble solution dominates the free energy for $T_\text{BTZ} < T_\star$, while the BTZ black string dominates for $T_\text{BTZ} > T_\star$.

We would like to better understand the physics of each phase of the field theory on BTZ black holes.  Let us start with the high $T_\text{BTZ}$ phase dual to the 4-dimensional BTZ black string.  Since the bulk metric takes the form \req{adsinads} (see previous footnote) the large $N$ stress tensor vanishes, i.e., it is of ${\cal O}(1)$ in the large $N$ limit.  Furthermore, even at finite $N$ the $T_\text{BTZ} \to \infty$ stress tensor is just the double Wick rotation of that on pure \AdS{3}.  Since \AdS{3} is conformally flat (and since there is no $d=3$ conformal anomaly), this vanishes exactly in any conformally-invariant renormalization scheme. Further analysis of the BTZ black string phase is similar to that of the \SAdS{} black string discussed in \sec{s:sads}.  The bulk solution is again a black funnel in the sense described in \sec{s:sads}, and the $q{\bar q }$ potential in the field theory displays no linear regime.

However, the bubble-of-nothing phases are quite different. At large $N$, we may compute the stress tensor either by double Wick rotation of known results for phases on \AdS{3} or by a holographic computation of the boundary stress tensors for the relevant 4-dimensional bulk duals \req{sadsbtzF} with $f(\rho)$ given as in \req{fsads4}.  We find
\begin{equation}
\CT_\mu^{\;\nu} = c\, \frac{\mu}{L \, \ell^3} \; \frac{r_+^3 }{3\, r^3}\, \big\{1,1,-2\big\}
\label{btzst}
\end{equation}	
to leading order in large $N$,  with $c \sim \CO({N^\frac{3}{2}})$. For the M2-brane world-volume theory the central charge $c$ is given in footnote \ref{M2c}.  Furthermore, $\mu = \mu(r_+)$ is defined by \req{fsads4} and \req{dcsads4}:
\begin{equation}
\mu(r_+) = \frac{4 \,  L \, \ell^3}{27 \, r_+^3} \left[1+\(1+\frac{3 \, r_+^2}{2 \, \ell^2} \right) \,
\sqrt{1-\frac{3 \, r_+^2}{\ell^2}}  \right] \ .
\label{}
\end{equation}	
As noted in \sec{s:ftads}, (\ref{btzst}) describes a negative energy density in parallel with known results \cite{Steif:1993zv} for free fields.

This result might suggest that the phase is deconfined.  Recall, however,  that in the limit $\rho_+ \to \infty$ the large \SAdS{4} black hole becomes the (unboosted) planar hole \req{poinbh}.  As a result, in the limit $T_\text{BTZ} \to 0$ the large \AdS{4} bubble of nothing obtained by double Wick rotation becomes just the \AdS{4} soliton,\footnote{This is the same \AdS{4} soliton spacetime which appeared as a boundary spacetime in \sec{s:sads}, where it played the role of a background for a 4-dimensional field theory in a context where gravity was not dynamical.  In contrast, here the 4-dimensional soliton approximates the bulk \AdS{4} dual of a state of a 3-dimensional CFT on the BTZ background. In particular, here the 4-dimensional soliton arises as the solution of the bulk gravitational equations of motion.} a static solution which is the prototypical example of the bulk dual to a confining phase \cite{Witten:1998zw}!\footnote{The phase transition for field theories with Scherk-Schwarz boundary conditions was also discussed in \cite{Myers:1999psa,Surya:2001vj}, which are all similar to the Hawking-Page transition originally described in   \cite{Hawking:1982dh}.} As is well known, there is no tension between the confining properties of this state and its ${\cal O}(N^{3/2})$ stress tensor.  For $T_{BTZ}$ small but non-zero, the bubble of nothing spacetimes have much in common with those discussed\footnote{Or, at least, the analogous solutions obtained by removing the so-called UV brane.} in \cite{Fitzpatrick:2006cd} in which the horizon extends from the boundary down to an AdS-soliton-like IR floor.

The \AdS{4} soliton  of radius $2\pi \, \ell  L \, T_\text{BTZ}$ and AdS length scale $L$
 is a useful approximation to the \AdS{4} bubbles of nothing in the region far from the bulk horizon.  There it leads immediately to a linear $q{\bar q}$  potential\footnote{The coefficient of this linear relation has a mild position dependence that varies on length scales of $\CO(\ell)$.} on scales much smaller than $\ell$ but much larger than $r_+$.

\subsubsection{Rotating BTZ black hole}
\label{s:kerrbtz}

Having understood the story for the static BTZ black hole, we can now turn to the more general case allowing rotation. Since much of the analysis is similar, we will be brief, emphasizing only the salient points of interest in the rotating case.

Our starting point will be the family of bulk asymptotically \AdS{4} geometries dual to stationary equilibrium configurations of CFTs on $\Sp^2\times \R$ with a non-zero chemical potential for rotation. The gravity duals of strongly coupled CFTs are believed to be the rotating-\AdS{4} black holes discovered in \cite{Carter:1968cz} whose metric in conventional Boyer-Lindquist coordinates takes the form:
\begin{eqnarray}
\label{KAdS4}
ds^2 &=& -\frac{\Delta_\rho}{\zeta^2} \, \(dT - \frac{a}{\Xi}\, \sin^2\Theta\, d\Phi\)^2 + \frac{\Delta_\Theta\, \sin^2\Theta}{\zeta^2} \, \(a\, dT - \frac{\rho^2 + a^2}{\Xi}\, d\Phi\)^2 \nonumber \\
 && \qquad + \; \frac{\zeta^2}{\Delta_\Theta}\, d\Theta^2 + \frac{\zeta^2}{\Delta_\rho}\, d\rho^2,
\label{kerrads}
\end{eqnarray}
with various metric functions given by
\begin{eqnarray}
\Delta_\rho(\rho) &=& (\rho^2+ a^2) \(1+ \frac{\rho^2}{L^2}\) - \mu \, \rho \ , \qquad \mu = \frac{1}{\rho_+}\, (\rho_+^2+ a^2) \(1+ \frac{\rho_+^2}{L^2}\)
\nonumber \\
\Delta_\Theta(\Theta) &=& 1-\frac{a^2}{L^2}\, \cos^2\Theta \ , \qquad \zeta^2(\rho,\Theta) = \rho^2 +a^2\, \cos^2\Theta \ , \qquad \Xi = 1-\frac{a^2}{L^2}.
\label{kerrfns}
\end{eqnarray}
This yields a two parameter family of solutions, with conserved charges $E = \mu/\Xi^2$ and $J = a\, \mu/\Xi^2$, corresponding to  the energy and angular momentum respectively.   They reduce to \SAdS{4} when $a= 0$, and to pure \AdS{4} when $\mu  =0$ (which is however expressed in rotating coordinates when $a\neq 0$). The solutions have a horizon (for a certain range of the parameters $\mu$ and $a$)  at the largest root of $\Delta_\rho$, which we denote as $\rho_+$. Note that in order for \req{kerrads} to describe a Lorentzian metric, the parameter $a$ is constrained to take values in a finite domain $0 \leq a^2 \leq L^2$.  (Without loss of generality we will take $0 \le a \le L$.)

These geometries have a conformal boundary which is a rotating \esu{3}. One can map this to a static \esu{3} by a coordinate transformation \cite{Hawking:1998kw} (see \req{kerresu} below) which allows one to read off the angular velocity of the rotating AdS black holes with respect to a static frame at infinity.  The result is
\begin{equation}
\Omega_{\text{AdS}_4}  = \frac{a}{L^2}\, \frac{ \rho_+^2 + L^2}{\rho_+^2+a^2} \ .
\label{omkerr}
\end{equation}
The solutions are expected to be stable for all $|\Omega_{\text{AdS}_4}\, L |\leq 1$ \cite{Hawking:1999dp}, as there exists a Killing field which is timelike everywhere outside the horizon.\footnote{In terms of the coordinates introduced in \req{kerresu} which makes the metric conformal to \esu{3}, this is the horizon generator
$\p_T + \Omega_{\text{AdS}_4}\, \p_{\tilde \Phi}$.}

We now want to apply the double Wick rotations described in \sec{s:ftads} to construct four dimensional geometries which have the rotating BTZ black holes as their conformal boundary. This is achieved by the two step coordinate transformation,
\begin{eqnarray}
y\,\cos \theta &=& \rho \,\cos \Theta  \ ,  \nonumber \\
y^2 &=&\frac{1}{\Xi}\, \left(\rho^2\,\Delta_\Theta + a^2 \, \sin^2 \Theta\right) ,\nonumber \\
 {\widetilde \Phi}&=&\Phi+\frac{a}{L^2} \, T \ ,
\label{kerresu}
\end{eqnarray}	
followed by
\begin{eqnarray}
T &=& - i \, L \left( \frac{r_-}{\ell^2} \, t - \frac{r_+}{\ell} \, \phi \right)  , \nonumber \\
{\widetilde \Phi} &=& - i \, \left( \frac{r_+}{\ell^2} \, t - \frac{r_-}{\ell} \, \phi \right)  ,\nonumber \\
\tan \theta &=& \sqrt{ \frac{r^2 - r_+^2}{r_+^2-r_-^2}} \ ,
\label{kerrbtzF}
\end{eqnarray}	
which one can check reduces to the one employed  in \sec{s:sadsbtz} when $r_-=0$, i.e., when the black holes on the boundary and bulk are static, cf.\ \req{wick} and \req{dStoBTZcx}.

The first of these transformations is the one described in \cite{Hawking:1998kw}, which has the effect of mapping the rotating \AdS{4} black hole \req{kerrads} to a frame where the conformal boundary is a static \esu{3}.  The second step then simultaneously implements a conformal transformation of the boundary to \AdS{3} and (the inverse of) the double Wick rotation \req{BTZcoordchange}.  As one can readily tell, the net coordinate transformation is quite messy and we will not write down the final bulk metric explicitly.

In writing the transformation \req{kerrbtzF} we have introduced two new parameters $r_\pm$. These are however not independent parameters but are rather determined by the parameters of the four dimensional spacetime \req{kerrads}. This can be understood as follows: the coordinates $(T,{\widetilde \Phi} )$ in the rotating \AdS{4} geometry have certain periodicity conditions to ensure regularity of the spacetime:
\begin{equation}
(T,{\widetilde \Phi} ) \to (T , {\widetilde \Phi} + 2\pi) \ , \qquad (T,{\widetilde \Phi} ) \to \left(T + \frac{i}{T_{\text{AdS}_4} } , {\widetilde \Phi}  + i\, \frac{\Omega_{\text{AdS}_4} }{T_{\text{AdS}_4} }\right) .
\label{}
\end{equation}	
For the rotating \AdS{4} black hole, the angular velocity $\Omega_{\text{AdS}_4}$ is given by \req{omkerr} and the temperature is
\begin{equation}
T_{\text{AdS}_4} = \frac{\rho_+ \,(1+a^2 \, L^{-2} + 3\, \rho_+^2\, L^{-2} - a^2\, \rho_+^{-2})}{4\pi\, (\rho_+^2 +a^2)}\ ,
\label{rpkerr}
\end{equation}
while for pure \AdS{4} (i.e.\ $\mu=0$), $T_{\text{AdS}_4}$ and $\Omega_{\text{AdS}_4}$ can be freely chosen. In either case, given $T_{\text{AdS}_4}$ and $\Omega_{\text{AdS}_4}$, the coordinate transformations \req{kerresu} and \req{kerrbtzF} then induce definite periodicities on the BTZ coordinates $(t,\phi)$.  Inverting the reasoning of section \sec{s:ftads} we find
\begin{equation}
T_\text{BTZ} = \frac{1-\Omega_\text{BTZ}^2 \, \ell^2}{4\pi^2\, T_{\text{AdS}_4}\,\ell\, L} \ , \qquad \Omega_\text{BTZ} = \frac{L}{\ell} \,\Omega_{\text{AdS}_4}
\label{tomrels}
\end{equation}	
where the BTZ temperature and angular velocity are related to $r_\pm$ as in \req{tempb}. In particular, for the solutions obtained from the rotating AdS black hole we find
\begin{eqnarray}
r_+ &=& \frac{2 \, \ell }{\rho_+ \,L}\; \frac{\rho_+^2+a^2}{1+a^2 \, L^{-2} + 3\, \rho_+^2\, L^{-2} - a^2\, \rho_+^{-2}} \ ,
 \nonumber \\
r_- &=& \frac{2\,a\, \ell }{L^2\,\rho_+} \;\frac{\rho_+^2+L^2}{1+a^2 \, L^{-2} + 3\, \rho_+^2\, L^{-2} - a^2\, \rho_+^{-2}}\ .
\label{rprmval}
\end{eqnarray}	
On the other hand, if we consider the transformations \req{kerresu} and  \req{kerrbtzF} applied to pure \AdS{4} by starting with \req{kerrads} with $\mu =0$, we may obtain any values of $r_+ \ge r_- \ge 0$ by choosing appropriate $T_{AdS} \ge 0$ and $1 \ge L \, \Omega_{AdS} \ge 0$. Alternately, one can pick $r_+$ arbitrarily and then fix  $r_-$ by matching  the angular velocities.

\begin{figure}[t]
\begin{center}
\includegraphics[scale=0.9]{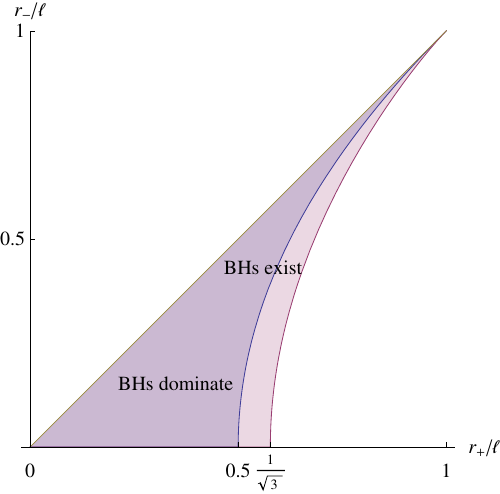}
 \caption{
  Regions in the $(r_+, r_-)$  plane where the bulk solution is allowed to be (the double Wick rotation of) a rotating-\AdS{4} black hole spacetime. Two such solutions are allowed for values of $r_\pm$  in the union of the two shaded regions, but none are allowed outside these regions. One of the (Wick rotated) black hole solutions has $\rho_+ > L$ only in the left shaded region. For boundary BTZ black holes lying in this region the dual bulk spacetime is the rotating-\AdS{4} solution \req{kerrads} to which the transformations \req{kerresu} and \req{kerrbtzF} have been applied.
  }
\label{f:rotbtz}
\end{center}
\end{figure}

\paragraph{The parameter space of rotating-\AdS{4} black holes:} We have parameterized rotating \AdS{4} black holes by two parameters, $\rho_+$ and $a$; however, physically sensible solutions exist only in part of this parameter space.  Furthermore, we are interested only in those \AdS{4} black holes for which the coordinate transformations \req{kerresu} and  \req{kerrbtzF} yield a 4-dimensional bulk, free of naked singularities, which induces a physically sensible BTZ black hole metric on the boundary.
The resulting constraints can be understood easily by examining the temperature $T_{\text{AdS}_4}$ and angular velocity $\Omega_{\text{AdS}_4}$.  For instance, the positivity of temperature \req{rpkerr}  requires $\rho_+ \geq \rho_+^\text{min}(a)$ where
\begin{equation}
\rho_+^\text{min}(a) = \frac{L}{\sqrt 6} \sqrt{- 1-\frac{a^2}{L^2}+\sqrt{1+14\, \frac{a^2}{L^2}+\frac{a^4}{L^4}}} \ .
\label{ecurve}
\end{equation}
This is the extremal locus for rotating \AdS{4} black holes.  In addition, noting that $\Omega_\text{BTZ} \, \ell = \Omega_{\text{AdS}_4}\, L$, the constraint $\Omega_{BTZ} \ell  \leq 1$ implies that
\begin{equation}
a \leq \text{min}\left(L, \;\frac{\rho_+^2}{L}\right).
\label{const2}
\end{equation}	
While for $\rho_+ > L$ this is just the constraint $a \le L$ required to ensure the correct signature of \req{kerrads}, for $\rho_+ < L$ it is in  fact stronger than both this signature constraint and the extremality constraint \req{ecurve}.  As a result of \req{const2}, one finds that at most two \AdS{4} black holes define bulk solutions consistent with a given BTZ metric on the boundary.  Furthermore, we comment that at most one of these black holes has $\rho_+ > L$.  The particular region in the $(r_+, r_-)$ plane where bulk black holes exist is shown in \fig{f:rotbtz}.  This can be determined analytically by eliminating $a/L$ from \req{rprmval} and then extremizing $\frac{r_-}{\ell}(\frac{r_+}{\ell},\frac{\rho_+}{L})$ over  $\frac{\rho_+}{L}$ for fixed $\frac{r_+}{\ell}$.

\begin{figure}[h]
\begin{center}
\includegraphics[scale=1]{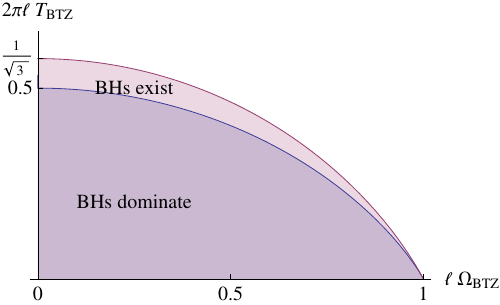}
 \caption{
  Regions in the  $(T_\text{BTZ}, \Omega_\text{BTZ})$ plane where the bulk solution is allowed to be (the double Wick rotation of) a rotating-\AdS{4} black hole spacetime.  As in \fig{f:rotbtz} two black hole solutions exist in the union of the shaded regions. In the lower of the shaded regions we have rotating \AdS{4} black holes with $\rho_+ > L$.  }
\label{f:rotbtzT}
\end{center}
\end{figure}

\paragraph{Thermodynamics:}  Recall that for any boundary BTZ black hole \req{BTZmet} with given values of $r_+$ and $r_-$, a Wick rotated vacuum \AdS{4} metric with appropriate identifications is always an allowed saddle point of the bulk gravity path integral.  In addition, we have the solutions constructed from \AdS{4} black holes as above.  Thus, for a given boundary BTZ black hole we have either:
(i) a single bulk solution which is  pure \AdS{4} or (ii)  three solutions, the additional two of which are analytic continuations of rotating \AdS{4} black holes with $\mu > 0$. The final question we need to address is the dominance of the saddle points. Which of the solutions at hand provides the correct bulk dual can be understood by looking at the  free energy difference between the rotating-\AdS{4} black holes and pure \AdS{4}. This  was computed carefully in \cite{Gibbons:2004ai} with the result
\begin{equation}
\Delta I = I_\text{rot-\AdS{4}} - I_\text{\AdS{4}} = -\frac{\pi}{\Xi\, L^4}\, \frac{(\rho_+^2 +a^2)^2 \,(\rho_+^2 -L^2)}{(3\,\rho_+^4 \, L^{-2} + (1+a^2\, L^{-2})\, \rho_+^2 - a^2)}.
\label{deltaI}
\end{equation}
In particular, from \req{deltaI} with the constraint \req{const2} we see that  $\Delta I <0$ if and only if $\rho_+  > L$; the rotating \AdS{4} black holes contribute when they are larger than the AdS scale. We can find
the boundary between the rotating \AdS{4} black holes and the pure \AdS{4} geometry for given values of $r_\pm$ by looking at the locus of points where $\rho_+ = L$ in \req{rprmval}. This region is also indicated in the  $r_\pm$ plane in \fig{f:rotbtz}, and is given by $r_+=\frac{r_-^2 + \ell^2}{2 \ell}$. A corresponding picture for the $(T_\text{BTZ}, \Omega_{\text{BTZ}})$ plane is shown in \fig{f:rotbtzT}.
In terms of  $(T_\text{BTZ}, \Omega_{\text{BTZ}})$, a bulk black hole dominates when
\begin{equation}
2 \pi \, \ell \, T_\text{BTZ}
 \le \frac{(1 - \Omega_{\text{BTZ}}^2 \, \ell ^2)\, \left( 1 - \sqrt{1 - \Omega_{\text{BTZ}}^2 \, \ell ^2} \right) }{\Omega_{\text{BTZ}}^2 \, \ell^2} \ ;
\label{Tdom}
\end{equation}	
i.e., for small $T_{BTZ}$ in parallel with the static case discussed earlier.


\paragraph{The quantum stress tensor in rotating BTZ background:}
An interesting question is the nature of the expectation value of the quantum stress tensor for a strongly coupled theory on the rotating  BTZ black hole background. One could directly compute this by using the coordinate transformations given above \req{kerresu}, \req{kerrbtzF}, but that is rather cumbersome given the nature of the coordinate transformations. Rather, we will use the fact that \req{kerrbtzF} is a bulk coordinate transformation that acts as a boundary conformal mapping. In particular, it maps the static \esu{3} with metric
\begin{equation}
ds^2  = -dT^2 + L^2 \, \left(d\theta^2 + \sin^2\theta\, d{\widetilde \Phi}^2 \right)
\label{esumet}
\end{equation}	
to the BTZ background \req{BTZmet} up to a conformal factor
\begin{equation}
e^{2 \,\omega} =  \frac{L^2 \, (r_+^2-r_-^2)}{\ell^2\,(r^2-r_-^2)}.
\label{conffactor}
\end{equation}	
This means that we can take the known stress tensor for the rotating AdS solution $(\CT_{\text{AdS}_4})^{\mu\nu}$ and use the coordinate change \req{kerrbtzF} together with the conformal transformation \req{conffactor} to obtain the boundary stress tensor on the rotating BTZ background. We have in particular,
\begin{equation}
\CT^{\mu\nu} = e^{5\omega} \, (\CT_{\text{AdS}_4})^{\mu\nu} \ .
\label{}
\end{equation}	

The stress tensor of the rotating \AdS{4} geometry \req{kerrads} with conformal boundary taken to be the non-rotating \esu{3} is given as \cite{Bhattacharyya:2007vs}:
\begin{eqnarray}
(\CT_{\text{AdS}_4})^{TT} &=& \frac{\mu}{16\pi\, G_N^{(4)}}\, \frac{1}{L^2}\, \gamma^5 \, (3 - \gamma^{-2}) \nonumber \\
(\CT_{\text{AdS}_4})^{T{\widetilde \Phi}} &=&3\,  \frac{\mu}{16\pi\, G_N^{(4)}}\,\frac{1}{L^3}\, \gamma^5  \,\hat{\Omega}\nonumber \\
(\CT_{\text{AdS}_4})^{{\widetilde \Phi} {\widetilde \Phi}} &=&  \frac{\mu}{16\pi\, G_N^{(4)}}\,\frac{1}{L^4}\,\gamma^5  \,(3 \, \hat{\Omega}^2 \,L^2+ \gamma^{-2}\,\csc^2\theta) \nonumber \\
(\CT_{\text{AdS}_4})^{\theta\theta}&=& \frac{\mu}{16\pi\, G_N^{(4)}}\, \frac{1}{L^4}\,\gamma^3
\label{kerrTmunu}
\end{eqnarray}	
where
\begin{equation}
\gamma = \frac{1}{\sqrt{1-\hat{\Omega}^2\,L^2\, \sin^2\theta}} \,, \qquad \hat{\Omega} = \frac{a}{L^2} \,.
\label{}
\end{equation}	
Note that we have written the stress tensor adapted to the \esu{3} metric given in \req{esumet}. We can use the holographic relation \req{centralch} to express the answer for the stress tensor in terms of the central charge of the field theory $c$.

Implementing the transformations described above we then obtain the following stress tensor on the rotating BTZ background in the coordinates $(t,r,\phi)$ of \req{BTZmet}:
\begin{equation}
\CT^a_{\ \; b} = c\, \frac{\mu}{L\, \ell^3}\; \frac{r_+^3}{r^3}\,
\left(\begin{array}{ccc}1 & 0 & 0 \\0 & 1 & 0 \\3 \frac{r_+ \,r_-}{\ell\, r^2} & 0 & -2\end{array}\right) ,
\label{TrotBTZ}
\end{equation}	
where $c$ is the central charge. For the maximally supersymmetric M2 brane world-volume theory, $c$ is given in footnote \ref{M2c}. Note that the schematic form of the quantum stress tensor agrees qualitatively with the result of \cite{Steif:1993zv}.

\noindent
\emph{Note added (July 30, 2020)}: The stress tensor for Kerr-\AdS{4} was inaccurately reported in  \eqref{kerrTmunu}  in earlier versions, resulting in an error for the stress tensor on the rotating BTZ geometry \eqref{TrotBTZ}. The mistake made was in identifying the parameter $\hat{\Omega}$ with the angular velocity of the Kerr-\AdS{4} black hole, $\Omega_{\text{AdS}_4}$, in \eqref{omkerr}.  This  is accurate for large Kerr-\AdS{4} black holes, $\rho_+ \gg L$, but not otherwise. One can check by direct computation using standard AdS/CFT dictionary that the expression in terms of $\hat{\Omega}$ is indeed the correct conserved boundary stress tensor for the rotating AdS black hole on \esu{4}.  We have fixed the result for the Kerr-\AdS{4} stress tensor \eqref{kerrTmunu}, but have left the stress tensor for the BTZ geometry in \eqref{TrotBTZ} unchanged. Consequently, the expression in \eqref{TrotBTZ} is accurate only when $\rho_+\gg L$, whence from \eqref{rprmval} we have $r_+ = \frac{2\,\ell\,L}{3\,\rho_+}$ and $r_- = \frac{2 \,a\,\ell L}{3\, \rho_+}$. Expressions for the general case are available in \cite{Emparan:2020znc}. We thank Roberto Emparan, Antonia Frassino and Benson Way for alerting us to this error and for clarifying discussions. We would also thank R.~Loganayagam for a discussion on the Kerr-\AdS{4} stress tensor computation reported in \cite{Bhattacharyya:2007vs}.

\section{Discussion}
\label{s:discuss}

Our main point was to use double Wick rotation to relate Hartle-Hawking-like states on certain asymptotically AdS black holes to thermal states on horizon-free backgrounds.  In particular, states on BTZ black holes are related in this way to thermal states on \AdS{3}, and states on planar AdS black holes are related to thermal states on AdS soliton.  States on {\em high} temperature black holes are mapped to {\em low} temperature states on the horizon-free backgrounds with the same chemical potential $\Omega$.   As a result, thermodynamic quantities have well-defined limits as $T_\text{BTZ}, T_\text{planar} \to \infty$ determined by the \AdS{3} and AdS soliton ground states. For large $N$ field theories with phase transitions, the fact that one expects a unique low-temperature phase on pure \AdS{3} or the AdS soliton suggests that there is also a unique phase on high temperature BTZ or planar black holes.

Any finite temperature corrections to the above limit are also determined by low-temperature physics on \AdS{3} and the AdS soliton.  For free theories, this means that the corrections are exponentially small.  For confining theories at large $N$, the corrections vanish at leading order in the large $N$ expansion.

We also argued that double Wick rotation provides insight into Hartle-Hawking-like states on \SAdS{} black holes.  This provided a partial resolution of a puzzle raised in \cite{Gregory:2008br}, which found ${\cal O}(1)$ stress tensors on high-temperature black holes for 3+1 ${\cal N}=4$ SYM.  The resolution is partial in the sense that the double Wick rotated phase is a mild deformation of a confining phase on the AdS soliton background, where the theory has an ${\cal O}(1)$ stress tensor.  This confinement may provide a mechanism to ensure that the stress tensor remains ${\cal O}(1)$ under the above deformations, though we were not able to complete this argument.

A second part of our work above used AdS/CFT to explore in detail certain example phases of strongly-coupled theories on asymptotically AdS black hole backgrounds.  In particular, we studied  the phase described in \cite{Gregory:2008br}  for ${\cal N}=4$ SYM (dual to the \SAdS{}-black string) and a similar high-temperature phase of the M2-brane theory on BTZ black holes. This allowed us to contrast the above double Wick rotation mechanism with other situations \cite{Fitzpatrick:2006cd,Hubeny:2009hr} where one may find small stress tensors on high temperature black hole backgrounds.  More specifically, despite the small stress tensors these examples showed no evidence of a confining quark/anti-quark potential on any length scale.  In addition, field theory excitations coupled strongly to the black hole in a manner similar to that found in the black funnel states of \cite{Hubeny:2009hr,Hubeny:2009ta}. We also showed that a similar phase  exists for M2 brane world-volume theory on BTZ backgrunds, though it dominates only at high $T_\text{BTZ}$.  In contrast, on low-temperature BTZ backgrounds we found both a  ${\cal O}(N^{3/2})$ response of the stress tensor to changes in the metric and a linear quark/anti-quark potential over a large range of distance scales.

Though we did not study them explicitly in this work, it is clear that similar remarks also hold for certain higher-dimensional backgrounds.  In particular, BTZ black strings are double Wick rotations of \AdS{4} and similarly BTZ black $p$-branes are double Wick rotations of \AdS{p+3}.  Furthermore, when considering conformal field theories, one may use similar techniques (up to global issues) on any conformally flat black hole background. For example, by writing $\Sp^{d-1}$ in terms a warped product of $\Sp^2$ and $\Sp^{d-3}$, e.g.\ using the metric
\begin{equation}
ds^2_{\Sp^{d}} = d\theta^2 + \sin^2 \theta \,d \Phi^2 + \cos^2 \theta \,d \Omega^2_{d-3},
\end{equation}
one may use the sequence of Wick rotations (on $T,\Phi$) and conformal transformations described in \sec{s:btz} to map the $d$-dimensional Einstein static universe ($\mathbb{R} \times \Sp^{d-1}$) to BTZ$\times \Sp^{d-3}$. This allows one to consider a large class of CFTs on BTZ black hole backgrounds.  At least in strongly coupled planar limits, we expect the resulting phase structure to be similar to that described here for the M2 brane world-volume theories.

The examples studied in this work show that Hartle-Hawking-like states on asymptotically AdS black holes can simultaneously display behaviors commonly associated with both confined and deconfined phases of the given field theory.  As noted in \sec{s:adsconf}, even at large $N$, it appears difficult to give a useful definition of confined vs.\ deconfined phases on asymptotically AdS  black hole backgrounds.  However, it would be interesting to pursue this issue further.  In particular, let us again consider a case where the stress tensor is small due to confinement in the double Wick rotated spacetime.  In that case, one might like to understand if the theory admits some analogue of a Boulware vacuum state with large {\em negative} energy, so that the small Hartle-Hawking stress tensor could then be understood as a precise cancellation of this negative vacuum energy against a large positive contribution from a deconfined plasma.  In contrast, an analogue of a Boulware vacuum where the stress tensor remains ${\cal O}(1)$ would suggest that a deeper notion of confinement remains to be found.

\subsection*{Acknowledgements}

We thank Gary Horowitz, David Tong and  Diego Trancanelli for useful discussions and Simon Ross for comments on the manuscript. VEH and MR would like to thank the KITP for wonderful hospitality during the workshop ``Fundamental Aspects of Superstring Theory", as well as the  Pedro Pascual Benasque Center of Science and the Aspen Center for Physics for excellent hospitality during the course of this project. In addition, VEH, DM and MR would like to thank the ICTS, TIFR for hospitality during the Monsoon workshop in string theory where this project was initiated. VEH and MR are supported in part by STFC Rolling grant and by the US National Science Foundation under the Grant No.\ NSF PHY05-51164.  DM was supported in part by the US National Science Foundation under grants   PHY05-55669 and PHY08-55415 and by funds from the University of California.

\providecommand{\href}[2]{#2}\begingroup\raggedright\endgroup

\end{document}